\documentclass{aa}  

\usepackage{graphicx}
\usepackage{array}
\usepackage{txfonts}
\usepackage{lipsum}
\usepackage{hyperref}
\usepackage{url}
\usepackage{breqn}
\usepackage{diagbox}
\usepackage{cuted} 
\usepackage[mathscr]{euscript}
\usepackage[export]{adjustbox}
\usepackage{multirow}
\usepackage[nice]{nicefrac}
\usepackage{enumitem}
\usepackage{booktabs}
\usepackage{placeins}
\usepackage{capt-of}

\newcolumntype{C}{>{\centering\arraybackslash}p{0.2\textwidth}}

\usepackage{natbib}
\bibpunct{(}{)}{;}{a}{}{,} 
\begin{document} 
\title{The radiative torque spin-up efficiency of ballistic dust-grain aggregates}
\author{Jonathan A. Jäger \inst{ \inst{\ref{inst1},\ref{inst2}}} \and  Stefan Reissl \inst{\ref{inst1}} \and Ralf S. Klessen \inst{ \inst{\ref{inst1},\ref{inst3},\ref{inst4},\ref{inst5}}}
}

\institute{
\centering Universit\"{a}t Heidelberg, Zentrum f\"{u}r Astronomie, Institut f\"{u}r Theoretische Astrophysik, Albert-Ueberle-Straße 2, \\  69120 Heidelberg, Germany \label{inst1} 
\and
\centering Max-Planck-Institut für Astrophysik, Karl-Schwarzschild-Str. 1,
D-85748 Garching, Germany  \label{inst2}
\and
\centering Universit\"{a}t Heidelberg, Interdisziplin\"{a}res Zentrum f\"{u}r Wissenschaftliches Rechnen, Im Neuenheimer Feld 205, \\ D-69120 Heidelberg, Germany  \label{inst3}
\and
\centering Harvard-Smithsonian Center for Astrophysics, 60 Garden Street,\\Cambridge, MA 02138, U.S.A. \label{inst4}
\and
\centering Elizabeth S. and Richard M. Cashin Fellow at the Radcliffe Institute for Advanced Studies at Harvard University, 10 Garden Street,\\Cambridge, MA 02138, U.S.A. \label{inst5}
}
						
\abstract
   {}
    {It is quintessential for the analysis of the observed dust polarization signal to understand the rotational dynamics of interstellar dust grains. Additionally, high rotation velocities may rotationally disrupt the grains, which impacts the grain-size distribution. We aim to constrain the set of parameters for an accurate description of the rotational spin-up process of ballistic dust grain aggregates driven by radiative torques (RATs).}
   {We modeled the dust grains as complex fractal aggregates grown by the ballistic aggregation of uniform spherical particles (monomers) of different sizes. A broad variation of dust materials, shapes, and sizes were studied in the presence of different radiation sources.}
   {We find that the canonical parameterization for the torque efficiency overestimates the maximum angular velocity $\omega_{\mathrm{RAT}}$ caused by RATs acting on ballistic grain aggregates. To resolve this problem, we propose a new parameterization that predicts $\omega_{\mathrm{RAT}}$ more accurately. We find that RATs are most efficient for larger grains with a lower monomer density. This manifests itself as a size- and monomer-density dependence in the constant part of the parameterization. Following the constant part, the parameterization has two power laws with different slopes that retain universality for all grain sizes.
   The maximum grain rotation does not scale linearly with radiation strength because different drag mechanisms dominate, depending on the grain material and environment. The angular velocity $\omega_{\mathrm{RAT}}$ of individual single dust grains has a wide distribution and may even differ from the mean by up to two orders of magnitude. Even though ballistic aggregates have a lower RAT efficiency, strong sources of radiation (stronger than $\approx 100$ times the typical interstellar radiation field) may still produce rotation velocities high enough to cause the rotational disruption of dust grains.}
 {}
  \maketitle

\section{Introduction}
Dust is a common component in astronomic environments from galaxies to stellar disks. It therefore plays a crucial role in observations. Most astronomers come into contact with dust because of dust extinction, which affects almost all observations \citep{draine_interstellar_2003}. \cite{calzetti_dust_2000} and \cite{bernstein_first_2002} estimated that $30\ \%$ or more of the energy that is produced by stars in the Universe is absorbed by dust and is reemitted in the infrared. 
The dust helps to cool the gas, which facilitates gravitational collapse \citep{omukai_thermal_2005} and also plays a major role in later stages of star formation. It also constitutes the first stage in the formation of planets \citep{dullemond_dust_2008, liu_tale_2020}. Rapidly rotating dust grains are the common source of polarized light and carry information about the magnetic field orientation \citep[see e.g.][for review]{Lazarian2007JQSRT,2015ARA&A..53..501A}. The catalysis of molecular hydrogen \citep{hollenbach_surface_1971, cazaux_molecular_2002} and other astrochemical reactions in the interstellar medium (ISM) \citep{dishoeck_astrochemistry_2014} is also impacted by grain rotation \citep{Hoang2020ApJ}. 
The first models trying to explain the observed dust alignment, that is, paramagnetic relaxation \citep{davis_polarization_1951} or the Gold mechanism \citep{gold_polarization_1952}, both encountered problems and were ultimately unable to account for the observed alignment. A great leap in the study of alignment was made when \cite{purcell_suprathermal_1979} realized that grains could experience systematic torques acting in a specific direction in the grain coordinates for long time spans compared to the timescale that it takes gas collisions to randomize the grain rotation. With these systematic torques, the grain can spin up to much faster angular velocities than the typical thermal rotation rate.
With suprathermal rotation, a grain can remain in a given orientation for much longer than the gas-collision timescale. This allows an alignment of the angular momentum with the maximum moment of inertia via inelastic relaxation or the Barnett relaxation, which was discovered by \cite{purcell_suprathermal_1979} and is based on the \cite{barnett_magnetization_1917} effect, which causes magnetization along the rotation axis in a paramagnetic body. When the grain rotation axis is not aligned with a principal axis of inertia, the rotation axis and therefore the magnetization axis will precess in body coordinates, dissipating energy and aligning the rotation axis with the axis of the maximum moment of inertia.
In the presence of an external magnetic field, the Davis-Greenstein process can align the grain with the magnetic field \citep{purcell_suprathermal_1979}.

A one-to-one mapping of dust alignment to the magnetic field lines would be of great use to the study of magnetic fields. Polarization observations of the object in question could then always be used to obtain information about the magnetic field morphology. Uncertainties about the completeness of the model remained, however, such as the lack of an increase in polarization with extinction in cold dark clouds \citep{goodman_does_1995} and the alignment of grains in cometary dust \citep{harwit_alignment_1971}. The exploration of other alignment mechanisms therefore continued. 

The idea of systematic radiative torques (RATs) as the cause for rapid grain rotation had already been advanced decades before
\citep{dolginov_orientation_1972,dolginov_orientation_1976},
but it only became an integral part of dust alignment theory with the work of \cite{DraineWeingartner1996}. Using the discrete dipole approximation (DDA) method, the authors showed that scattering and absorption of light from irregular grains results in torques that act on timescales that are much longer than random gas collisions.
The calculations were made on a irregular shape of intersecting cuboids, and the authors found that radiative torques strongly depend on the grain size. Additional grain shapes were investigated in the following work \citep{DraineWeingartner1997} in the context of anisotropic radiation, $\mathrm{H}_2$-formation torques, and paramagnetic relaxation. In most cases, the grain reached long-term stable conditions by aligning the principal grain axis with the magnetic field. \cite{lazarian_radiative_2007} (LH07 hereafter) later introduced an analytic model (AMO) for a spherical grain that is attached to a flat rectangular mirror via a weightless noninteracting rod. In this way, they captured the helical properties of irregular grains. The AMO was compared to DDA simulations of irregular grain models that corresponded to the grain shapes in \cite{DraineWeingartner1997}, and a good agreement was found. The alignment of the principal grain axis with the direction of incoming radiation was confirmed when magnetic field effects were subdominant. The parameterization of the RAT spin-up efficiency was found to be a power law with an exponent of about $-3$.  Physical implications of RATs were explored in following studies \citep{LazarianHoang2008,hoang_radiative_2008,hoang_radiative_2009,2016ApJ...831..159H}, and the model was expanded by the addition of superparamagnetic relaxation due to iron inclusions. This increased the range of parameters in which grains can efficiently align with the magnetic field, which is necessary to explain high degrees of polarization from observations \citep{planck_collaboration_planck_2020}.
The AMO was used successfully to model synthetic dust polarization observations as long as free alignment parameters could be used to adjust the model \citep{Bethell2007ApJ,hoang_grain_2014, reissl_radiative_2016}.
\cite{herranen_radiative_2019} then made a step toward testing more realistic grain shapes: They used Gaussian ellipsoids, oblate spheroids, prolate spheroids, and spheres with random deformations and found a shallower slope of $-2.6$ for the power law for the RAT parameterization.
In this work, we take one more step toward more realistic grain shapes by simulating random typically fractal aggregate grains composed of differently sized monomers. We compare their RATs to previous results. The motivation is to model the grain growth ab initio using the current understanding that dust is originally formed in AGB stellar winds or supernovae \citep{todini_dust_2001,schneider_dust_2014,nanni_evolution_2013}. As the dust-forming materials in the winds cool down, heavy elements condense into spherical particles  (in the following called monomers), which are the building blocks for the larger dust grains that are formed by monomer accretion and coagulation of grains \citep{gobrecht_dust_2016,norris_close_2012}. A more realistic grain modeling may result in differing interactions with radiation. One example was given by \cite{arnold_effect_2019}, who found significant differences in dust blowout sizes between compact and agglomerated grains.\\

The aim of this work is to improve the understanding of RATs by simulating individual grains composed of monomers for different sources of radiation. 
The properties of grains and the way they are modeled is presented in Sect. \ref{sect:modeling}. The theoretical basis for our simulations is presented in Sect. \ref{sect:RATRotation}, and models for the sources are described in Sect. \ref{sect:sources}. The simulation codes are explained in Sect. \ref{sect:DustProperties}, the results are shown in Sect. \ref{sect:results} and discussed in Sect. \ref{sect:discussion}, and a summary and outlook are given in Sect. \ref{sect:summary}.

\section{Ballistic dust grain aggregation} \label{sect:modeling}
We modeled dust grains as conglomerates consisting of spherical particles of a single material, called monomers. The monomers stay attached to each other due to the intermolecular forces at their contact area \citep[see e.g.][]{Johnson1971}. The growth of larger dust grains follows an aggregation process by means of ballistic collision and sticking of individual monomers \citep{Dominik1997}. To mimic the grain-growth process of ballistic aggregation (BA), we followed the procedure as outlined in \cite{Shen2008}, modified to work with monomers of variable sizes \citep{Reissl2023}. Starting with two or three connected monomers, new monomers with radii drawn from a log-normal distribution between $10\ \mathrm{nm}$ and $100\ \mathrm{nm}$ were injected into the simulation volume from random directions so that they stuck to the grain upon collision.

\begin{figure*}[htbp]
	\centering
	\includegraphics[width=0.33\textwidth]{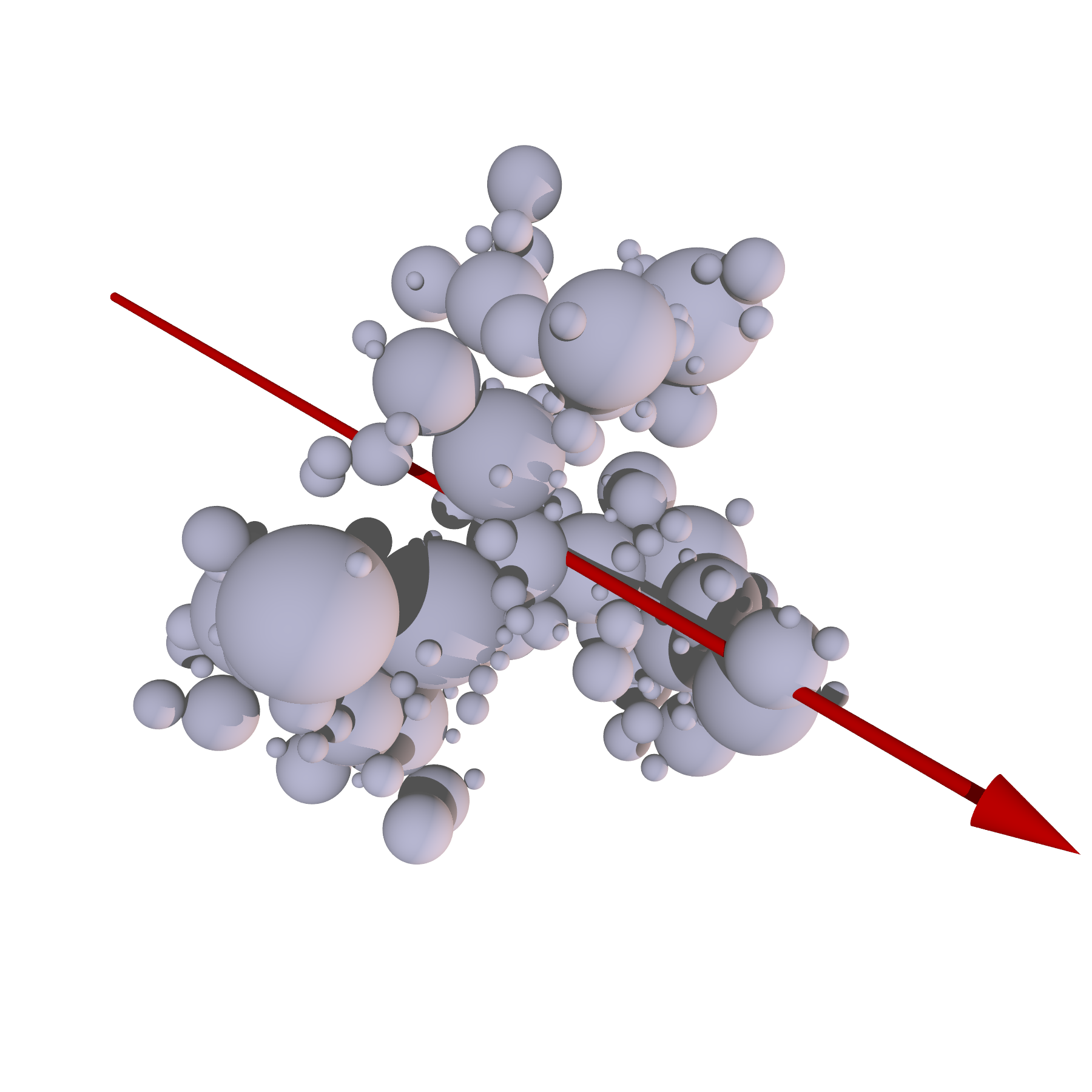}
    \includegraphics[width=0.33\textwidth]{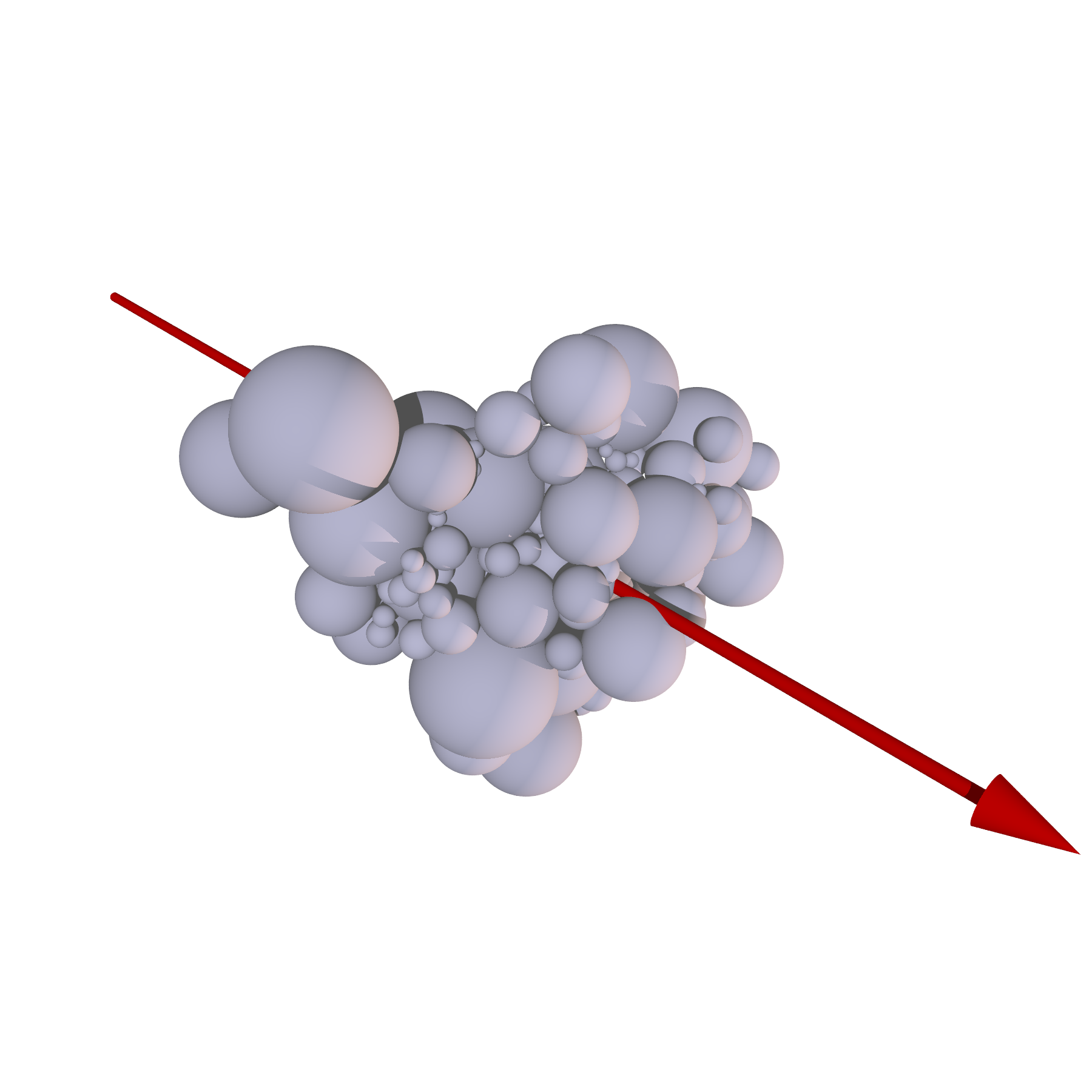}
    \includegraphics[width=0.33\textwidth]{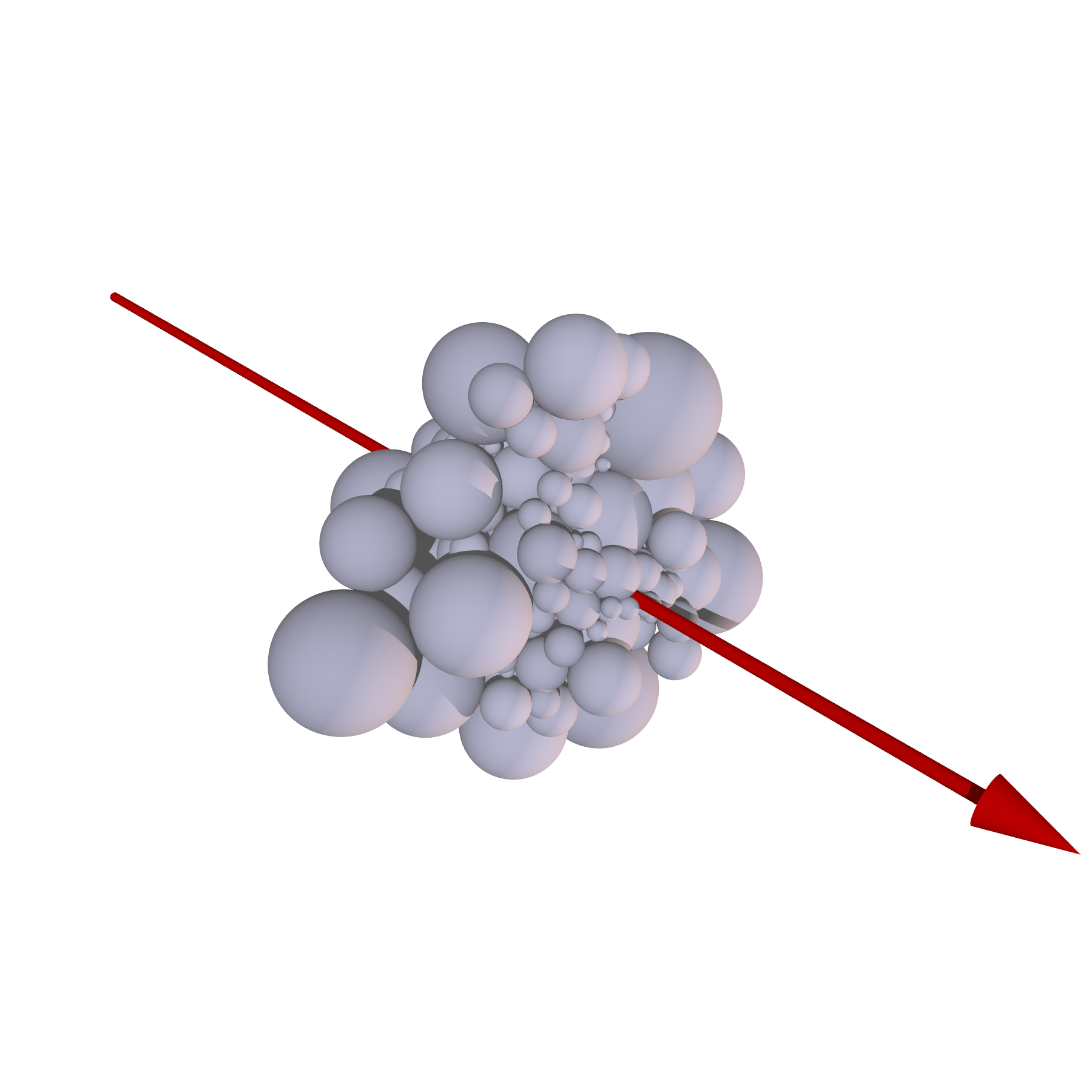}
	\caption{Exemplary  BA (left), BAM1 (middle), BAM2 (right) grains with an effective radius of ${ a_{\mathrm{eff}}=250\ \mathrm{nm} }$. The rotation  axis $\hat{a}_{\mathrm{1}}$ is depicted as the red arrow and is associated with the direction of the maximum moment of inertia $I_{\mathrm{max}}$. }
	\label{fig:Grains}
\end{figure*}

Grains with an effective radius $a_{\mathrm{eff}}$\footnote{The effective grain radius gives the radius of the grain as if it were reshaped into a perfect sphere.} from $50\, \mathrm{nm}$ up to $250\, \mathrm{nm}$ in steps of $50\, \mathrm{nm}$ were created. To reach an exact effective radius, the sizes of the final monomers added to the grain were sampled in a biased way (we lowered the upper bound on the monomer size), until the desired $a_{\mathrm{eff}}$ was reached exactly. In addition to BA, where each monomer remains at its initial sticking position, we modeled different monomer densities or porosities of aggregates by means of monomer migration. After hitting the forming grain, the added monomer rolls along the surface of the monomer it impacted until it touches a second monomer of the grain (BAM1). Here, all monomers are connected to at least two other monomers. For the BAM2 case each monomer rolls twice until it touches a third monomer. For BAM2, all monomers are connected to at least three other monomers. This is the densest way in which a random grain can be packed with monomers using the migration process. It follows that BA grains are the least  dense/most porous and BAM2 grains are the most dense/least porous, an exemplary grain of each monomer density is shown in Fig.~\ref{fig:Grains}.

To recapitulate, the three grain properties are the material from which the grain is made (astrosilicate or hydrogenated amorphous carbon), which differs in the optical properties; the monomer density of the grain (BA, BAM1, or BAM2), which determines how closely the monomers are packed; and finally, the grain size given by the effective radius ($50\, \mathrm{nm}$ to $250\, \mathrm{nm}$ in steps of $50\, \mathrm{nm}$). For every combination of these three properties, 30 different grains were created, leading to a total of 900 individual grains. This method of modeling ballistic grain aggregates was previously used in \cite{reissl_mechanical_2023}.

\section{Maximum grain rotation} \label{sect:RATRotation}
In the presence of a directed radiation field, irregularly shaped dust grains can experience momentum transfer from photon interactions. In general, the time-averaged forces acting on such a grain may result in a net torque, accelerating the rotation. This radiative torque (RAT) can be evaluated by \citep[see e.g.][and LH07]{DraineWeingartner1996}
\begin{equation}
\vec{\Gamma}_{\mathrm{RAT}} =  \frac{1}{2} \gamma
 a_{\mathrm{eff}}^2 \int \lambda  u_\lambda \vec{Q}_\Gamma   \,\mathrm{d}\lambda \,,
\label{eq:GammaRAT}
\end{equation}
where $\gamma$ describes the anisotropy of the radiation field, $\lambda$ is the wavelength, and the quantity $u_\lambda=F_\lambda/c$ is the spectral energy density of a radiation field with the spectral flux $F_\lambda$, whereas $c$ is the speed of light. These factors determine the overall strength of the torque, independent of the grain shape and orientation. \\The effect of the individual grain shape and material is contained within the dimensionless spin-up efficiency vector $\vec{Q}_\Gamma$ of the RAT. A perfectly spherical grain would experience no net torque at all because of the symmetry between collisions causing clockwise and anticlockwise torques. However, for grains with irregular shapes, the contributions of individual radiation-dust interactions lead to a nonzero $\vec{Q}_\Gamma$ vector, which modulates the orientation, that is, the sign as well as the magnitude of $\vec{\Gamma}_{\mathrm{RAT}}$.\\
In previous works, a power-law parameterization for $Q_\Gamma=|\vec{Q}_\Gamma|$ was suggested (LH07) 
\begin{equation}
 Q_\Gamma =\begin{cases} 0.4  &\mbox{if } \lambda \leq 1.8\,a_{\mathrm{eff}} \\ 
		    0.4 \left(\frac{\lambda}{1.8\, a_{\mathrm{eff}}  } \right)^{\alpha}  & \mbox{otherwise} \end{cases}\, .
 \label{eq:QCanonical}
\end{equation}

We refer to this relation as the canonical parameterization. Here, the constant part transitions into a power law at wavelengths equal to $1.8$ times the effective radius of the grain. The power-law exponent typically lies between $\alpha=-2.6$ \citep{herranen_radiative_2019} and $\alpha=-4$ (LH07).

For our grain dynamics modeling, we assumed a stable alignment of the grain rotation with the direction of incoming radiation.
For a single grain that stably rotates about its principal axis, the maximum possible angular momentum $\omega_{\mathrm{RAT}}$ caused by RATs can be calculated as
\begin{equation}
\omega_{\mathrm{RAT}} =  \frac{\tau_{\mathrm{drag}} \Gamma_{\mathrm{RAT}}}{ I_{\mathrm{max}} } \quad ,
 \label{eq:omegaRAT}
\end{equation}
where $\tau_{\mathrm{drag}}$ is the drag timescale, and $I_{\mathrm{max}}$ is the highest moment of inertia, defining the $a_1$-axis in the body-centric system of the grain \citep[see e.g.][as well as Fig.~ \ref{fig:Grains}]{DraineWeingartner1996}. 
The total drag acting on the grain is composed of the collisions with the surrounding gas and the loss of angular momentum by emitting infrared photons. The total drag timescale can be evaluated from the gas-drag timescale $\tau_{\mathrm{gas}}$ and photon-drag timescale $\tau_{\mathrm{IR}}$ with
\begin{equation}
\tau_{\mathrm{drag}} = \frac{1}{\tau_{\mathrm{gas}}^{-1} + \tau_{\mathrm{IR}}^{-1} },
\end{equation}
where a shorter timescale results in a stronger drag. Here, the gas-drag timescale 
\begin{equation}
\tau_{\mathrm{gas}} = \frac{I_{\mathrm{max}}}{n_{\mathrm{gas}} m_{\mathrm{gas}}  \varv_{\mathrm{th}}  a_{\mathrm{eff}}^4 Q_{\mathrm{gas}}} \quad ,
\label{eq:tau_gas}
\end{equation}
depends on the number density of the gas $n_{\mathrm{gas}}$ as well as on the gas mass $m_{\mathrm{gas}}$ per particle and the mean thermal velocity $\varv_{\mathrm{th}}$ of the gas \citep[see e.g.][]{DraineWeingartner1996}. The shape of the grain affects the gas drag, with some shapes increasing the ability of the gas to exert torque on the grain. This influence of the grain shape is quantified by the dimensionless gas-drag efficiency $Q_{\mathrm{gas}}$. We explored two environments with the same gas number densities $n_{\mathrm{gas}}$ and different temperatures $T_{\mathrm{gas}}$, as listed in Tab.~\ref{tab:conditions}. 
\begin{table}
\caption{Fiducial gas parameters for the considered environments}
\label{tab:conditions}
\centering
 \begin{tabular}{c c c} 
    \toprule
    & $n_{\mathrm{gas}}\ \mathrm{[m^{-3}]}$  &  $T_{\mathrm{gas}}\ \mathrm{[K]}$  \\
    \midrule
    ISM  & $5\cdot 10^7$  & $100$\\
    AGN  & $5\cdot10^7$ & $5\cdot10^{4}$ \\
    \bottomrule
\end{tabular}
\end{table}
 The photon-drag timescale 
\begin{equation}
\tau_{\mathrm{IR}} = \frac{c^2 I_{\mathrm{max}}}{  a_{\mathrm{eff}}^2  f_{\mathrm{IR}}( T_{\mathrm{dust}})} \quad 
\label{eq:tau_IR}
\end{equation}
depends on the integral over the photon emission,
\begin{equation}
f_{\mathrm{IR}}( T_{\mathrm{dust}} ) = 4\pi\int_{0}^{\infty} \lambda^2 B_{\lambda}\left( T_{\mathrm{dust}} \right) Q_{\mathrm{abs}}(\lambda)\, \mathrm{d}\lambda \, .
\label{eq:fIR}
\end{equation}
Here, $B_{\lambda}\left( T_{\mathrm{dust}} \right)$ is the Planck function modulated by the dimensionless efficiency of the absorption $Q_{\mathrm{abs}}(\lambda)$, and $T_{\mathrm{dust}}$ is the dust temperature. For simplicity, the dust is illuminated by a single source without extinction. For the dust heating, we assumed thermal equilibrium. Therefore, the temperature of a single dust grain can be evaluated via 
\begin{equation}
\int_{0}^{\infty} F_{\lambda} Q_{\mathrm{abs}}(\lambda)\,\mathrm{d}\lambda =4\pi\int_{0}^{\infty} B_{\lambda}\left( T_{\mathrm{dust}} \right)Q_{\mathrm{abs}}(\lambda)\, \mathrm{d}\lambda\,
\label{eq:Tdust}
\end{equation}
by equating the source flux $F_{\lambda}$ that is absorbed by the grain with the characteristic emitted flux $B_{\lambda}Q_{\mathrm{abs}}(\lambda)$ of each grain.
\begin{table}[htbp]
\caption{Parameters for radiation sources}
\label{tab:sources}
\centering
 \begin{tabular}{c c c c c c} 
 \toprule
   & $R\ \mathrm{[R_{\odot}]}$  &  $T_{\mathrm{eff}}\ \mathrm{[K]}$ & $L_{\mathrm{bol}}\ \mathrm{[L_{\odot}]}$ & $\gamma$ & $u\ [\mathrm{u_{ISRF}}]$\\
   \midrule
ISRF             & -  & - & - & 0.1 & 1.0\\
S1               & 50  & $1.5\cdot 10^{4}$ & $4.6\cdot 10^{5}$ & 0.7 & $5.6$\\
S2               & 80  & $2.5\cdot 10^{4}$ & $9.0\cdot 10^{6}$ & 0.9 & $1.1\cdot 10^{2}$\\
AGN              & -  & - & $10^{10}$ & 1.0 & $1.2\cdot 10^{5}$\\
 \bottomrule
\end{tabular}
\tablefoot{The table includes the stellar radius $R$ and the effective temperature $T_{\mathrm{eff}}$ for stellar sources S1 and S2, the bolometric luminosity $L_{\mathrm{bol}}$ for all localized sources, the anisotropy $\gamma$, and the radiation energy density $u$, given in relation to $u_{\mathrm{ISRF}} = 8.64\cdot 10^{-14}\ \mathrm{J\ m^{-3}}$. The distance to the localized sources S1, S2, and to the AGN is taken to be $10\, \mathrm{pc}$.}
\end{table}

\begin{figure}[htbp]
	\centering
	\includegraphics[width=0.49\textwidth,left]{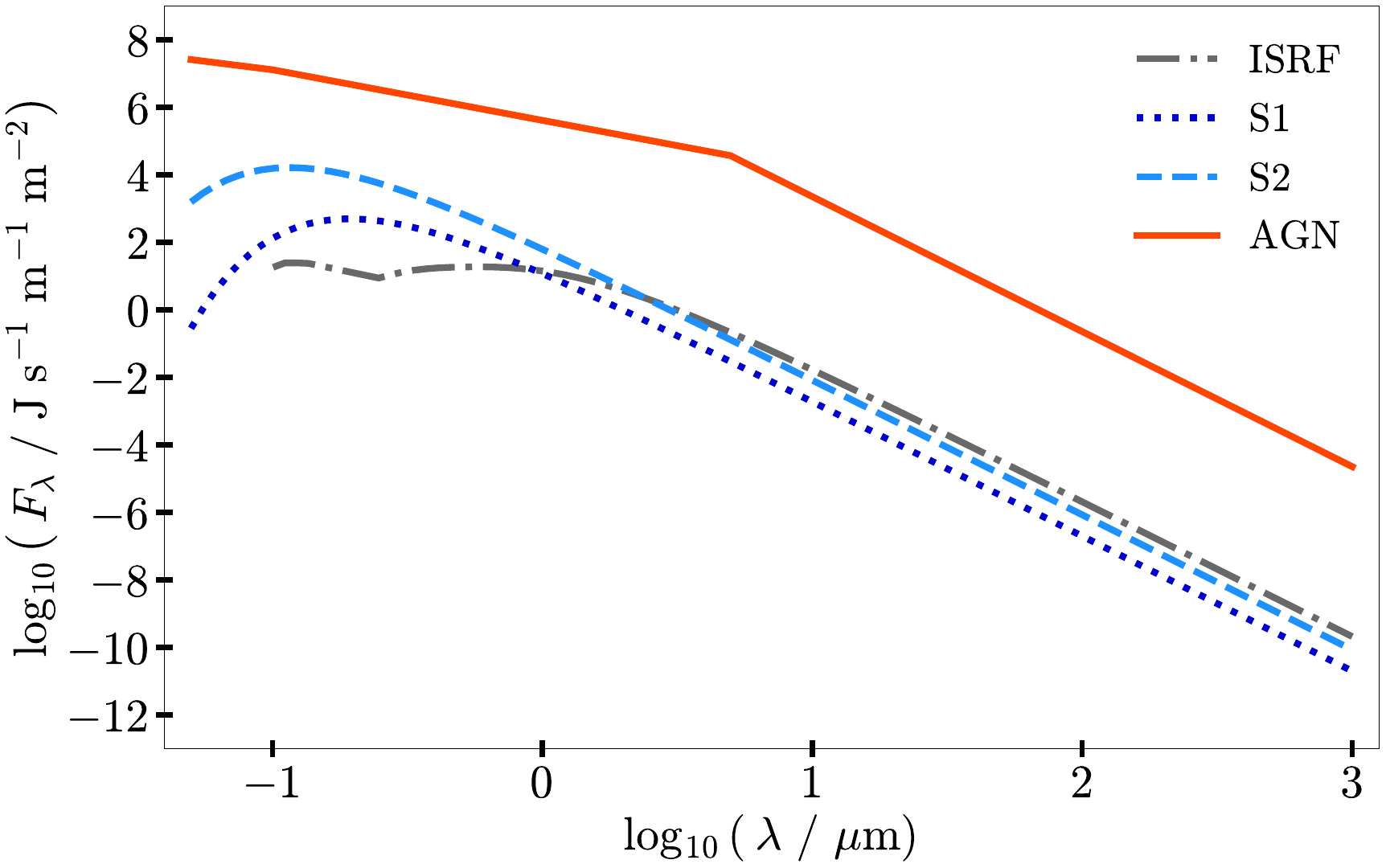}
	\caption{Spectral flux of sources ISRF, S1, S2, and AGN that were used to irradiate the grain aggregates.}
	\label{fig:sources}
\end{figure}

\section{The radiation field} \label{sect:sources}
Typical sources of radiation were chosen to evaluate the RATs they cause by irradiating the grains. The different sources included two O-type stars labeled S1 and S2, with a flux of 
\begin{equation}
F_\lambda= \pi \frac{R^2 }{d^2} B_{\lambda} \quad ,
\end{equation}
where $R$ is the stellar radius, and $d$ is the distance from the star ($1\, \mathrm{pc}$). The emission of an active galactic nucleus (AGN) was modeled as
\begin{equation}
    L^{\mathrm{AGN}}_\lambda= \frac{C}{\lambda} \cdot
    \begin{cases} 
        6.3 \cdot 10^{-4}  &\mbox{if } 0.01\, \mu\mathrm{m} <\lambda \leq 0.1 \, \mu\mathrm{m}\, , \\ 
		2 \cdot 10^{-4} \left(\frac{\lambda}{1 \mu\mathrm{m}} \right)^{-0.5}  & \mbox{if } 0.1 \, \mu\mathrm{m}<\lambda \leq 5 \, \mu\mathrm{m}\, , \\
        0.011 \left(\frac{\lambda}{1 \mu\mathrm{ m}} \right)^{-3}  & \mbox{otherwise}\, ,
    \end{cases}
\end{equation}
following \cite{AGN_model} and \cite{nenkova_agn_2008}, where $C$ is a constant chosen such that the integration of $L^{\mathrm{AGN}}_\lambda$ equals the typical AGN bolometric luminosity of ${ L_{\mathrm{bol}}=10^{10} L_{\odot} }$.\\
Finally, the spectral flux of the interstellar radiation field (ISRF) was parameterized as 

\begin{equation}
    F^{\mathrm{ISRF}}_\lambda= 4 \pi
    \begin{cases}
        0 &\mbox{if } \lambda \leq 0.0912 \, \mu\mathrm{m}\, ,\\ 
        3069 \left(\frac{\lambda}{1 \mu\mathrm{m}} \right)^{3.4172}  &\mbox{if } 0.0912\, \mu\mathrm{m} <\lambda \leq 0.11 \, \mu\mathrm{m}\, , \\ 
		1.627  & \mbox{if } 0.11 \, \mu\mathrm{m}<\lambda \leq 0.134 \, \mu\mathrm{m} \, ,\\
        0.0566 \left(\frac{\lambda}{1 \mu\mathrm{m}} \right)^{-1.6678}  & \mbox{if } 0.134 \, \mu\mathrm{m}<\lambda \leq 0.25 \, \mu\mathrm{m}\, , \\
        10^{-14}  B_{\lambda}\left( 7500 \mathrm{K} \right)+ \\  10^{-13} B_{\lambda}\left( 4000 \mathrm{K}\right)+ \\  4\cdot 10^{-13} B_{\lambda}\left( 3000 \mathrm{K}\right)  & \mbox{otherwise}\, ,
    \end{cases}
\end{equation}
following \cite{ISRF_model}, which is based on \cite{mathis_interstellar_1983} and \cite{weingartner_photoelectric_2001}.
The applied parameters of the radiation sources are listed in Tab.~\ref{tab:sources}, and the resulting spectral flux of the sources is shown in Fig.~\ref{fig:sources}. 

\section{Optical and mechanical dust properties} \label{sect:DustProperties}
To test the predictions of the RAT theory \citep[LH07; ][]{Hoang2014}, we considered silicates (SC) and carbonaceous (CA) grains. These materials are typically present in the ISM \citep[see e.g.][for review]{draine_interstellar_2003}. The optical properties of the materials are represented by the wavelength-dependent complex refractive index ${ \bar{n}=n+ik }$, with $n$ being the real part describing the refraction of light that interacts with the dust grain, and the imaginary part $k$ governs the light attenuation. For the SC, we took the (Astro)silicate model data as presented in  \cite{astrosil_Draine,laor_spectroscopic_1993}, and the  optical properties of hydrogenated amorphous carbon were outlined in \cite{jones_variations_2012} \footnote{
In detail: The refractive indices of the CA material were taken from \url{http://vizier.cds.unistra.fr/viz-bin/VizieR?-source=J/A+A/542/A98} for $30\ \mathrm{nm}$ hydrocarbon particles with $E_g = 2.67\,\mathrm{eV}$, and the SC data are listed as {\tt eps\_Sil} in \url{https://www.astro.princeton.edu/~draine/dust/dust.diel.html}}. Ice mantles were not considered here because high-intensity radiation environments would not allow them. Assuming condensation of all ice species at a temperature above $150\,\mathrm{K}$ \citep{zhang_evidence_2015} and comparing with dust temperatures in Tab.~\ref{tab:T_dust}, we would expect CA grains to be ice-free for the stellar sources as well as both CA and SC grains for the AGN source. The ice-free assumption may therefore generally not be valid for the ISRF and SC grains around the stellar sources. Additionally, a comparison with these dust temperatures assumes that there is no significant difference in the dust temperature when an icy surface is considered instead of the silicate or carbonaceous surface.

The radiative torque efficiency $\vec{Q}_\Gamma$ was calculated using the DDA code DDSCAT (version 7.3.3; see \citealp{draine_discrete-dipole_1994,draine_discrete-dipole_2008,flatau_fast_2012}).
The grain aggregates were initialized and tabulated as outlined in Sect.~\ref{sect:modeling}. To create input shape files for the DDSCAT simulations,  we filled the volume enclosed by the monomers with a cubic 3D grid, each grid point representing a distinct dipole, with properties determined by the complex refractive index.  
DDSCAT calculations of $\vec{Q}_\Gamma$ were performed in the external coordinate system ${ \{\hat{e}_1,\hat{e}_2,\hat{e}_3 \} }$ of incoming radiation, 
\begin{equation}
\vec{Q}_{\mathrm{\Gamma}}\left( \Theta \right) =  Q_{\mathrm{e1}}\left( \Theta \right)\hat{e}_{\mathrm{1}} + Q_{\mathrm{e2}}\left( \Theta \right)\hat{e}_{\mathrm{2}} + Q_{\mathrm{e3}}\left( \Theta \right)\hat{e}_{\mathrm{3}}\quad ,
\label{eq:Qcomponents}
\end{equation}
where $\hat{e}_{\mathrm{1}}$ is antiparallel to the wave vector of incoming light. For simplicity, we assumed that the grain was aligned with the direction of radiation, that is, $\Theta=0^\circ$. Subsequently, the direction of radiation and rotation coincide ($\hat{e}_{\mathrm{1}}\ ||\ \hat{a}_{\mathrm{1}}$), and the RAT efficiency factor becomes $\vec{Q}_{\mathrm{\Gamma}}\equiv Q_{\mathrm{e1}}\left( \Theta \right)\hat{e}_{\mathrm{1}}$. Since RATs are most efficient for $\Theta=0^\circ$, the data presented in this paper merely represent an upper limit for the modeling of the spin-up process of ballistic aggregates. However, the focus on the special case of $\Theta=0^\circ$ does not impact our comparison with previous attempts of modeling the efficiency of RATs. The simulation results were averaged over 21 equidistant values of the rotation angle of the grain around $\hat{a}_1$. The results were computed for 50 wavelengths between $0.05\, \mu\mathrm{m}$ and $150\, \mu\mathrm{m}$ within equidistant bins in log-space. Each combination of material, monomer density, size, and shape (see Sect.  \ref{sect:modeling}) was simulated with five different resolutions, that is, the separation of two nearest dipoles. They were combined using an average weighted by the dipole resolution. In addition to the  RAT efficiency factor  $\vec{Q}_{\mathrm{\Gamma}}$, DDSCAT also provides the efficiency of the absorption $\vec{Q}_{\mathrm{abs}}$ that is required to evaluate the photon-drag timescale $\tau_{\mathrm{IR}}$ and the dust temperature $T_{\mathrm{dust}}$. The gas-torque efficiency $Q_{\mathrm{gas}}$ needed to evaluate the gas-drag timescale $\tau_{\mathrm{gas}}$ was calculated with a Monte Carlo-based (MC) gas-dust collision approach as outlined in \cite{reissl_mechanical_2023}. The corresponding gas parameters\footnote{We note that the gas number density $n_{\mathrm{gas}}$ is lower than in a typical AGN torus. However, we can position the dust grain outside the torus, selecting a position in which $n_{\mathrm{gas}}$ is similar to the ISM, e.g., at higher inclinations from the plane of the torus.} are listed in Tab.~\ref{tab:conditions}.

\begin{table}[htbp]
\caption{Range and mean of individual grain dust temperatures}
\label{tab:T_dust}
\centering
 \begin{tabular}{l r r r}
 \toprule
 source & $T_{\mathrm{min}}$ [K] & $T_{\mathrm{mean}}$ [K] & $T_{\mathrm{max}}$ [K]
 \\
 \midrule
 \multicolumn{4}{c}{CA} \\
 \midrule
 AGN & 368 & 410 & 471\\
 S2 & 160 & 172 & 189\\
 S1 & 114 & 121 & 130\\
 ISRF & 72 & 79 & 88\\
 \midrule
 \multicolumn{4}{c}{SC} \\
 \midrule
 AGN & 119 & 136 & 161\\
 S2 & 37 & 41 & 48\\
 S1 & 22 & 24 & 26\\
 ISRF & 15 & 16 & 18\\
 \bottomrule
 \end{tabular}
\end{table}

\begin{figure*}[htbp]
	\centering
	\includegraphics[width=0.33\textwidth]{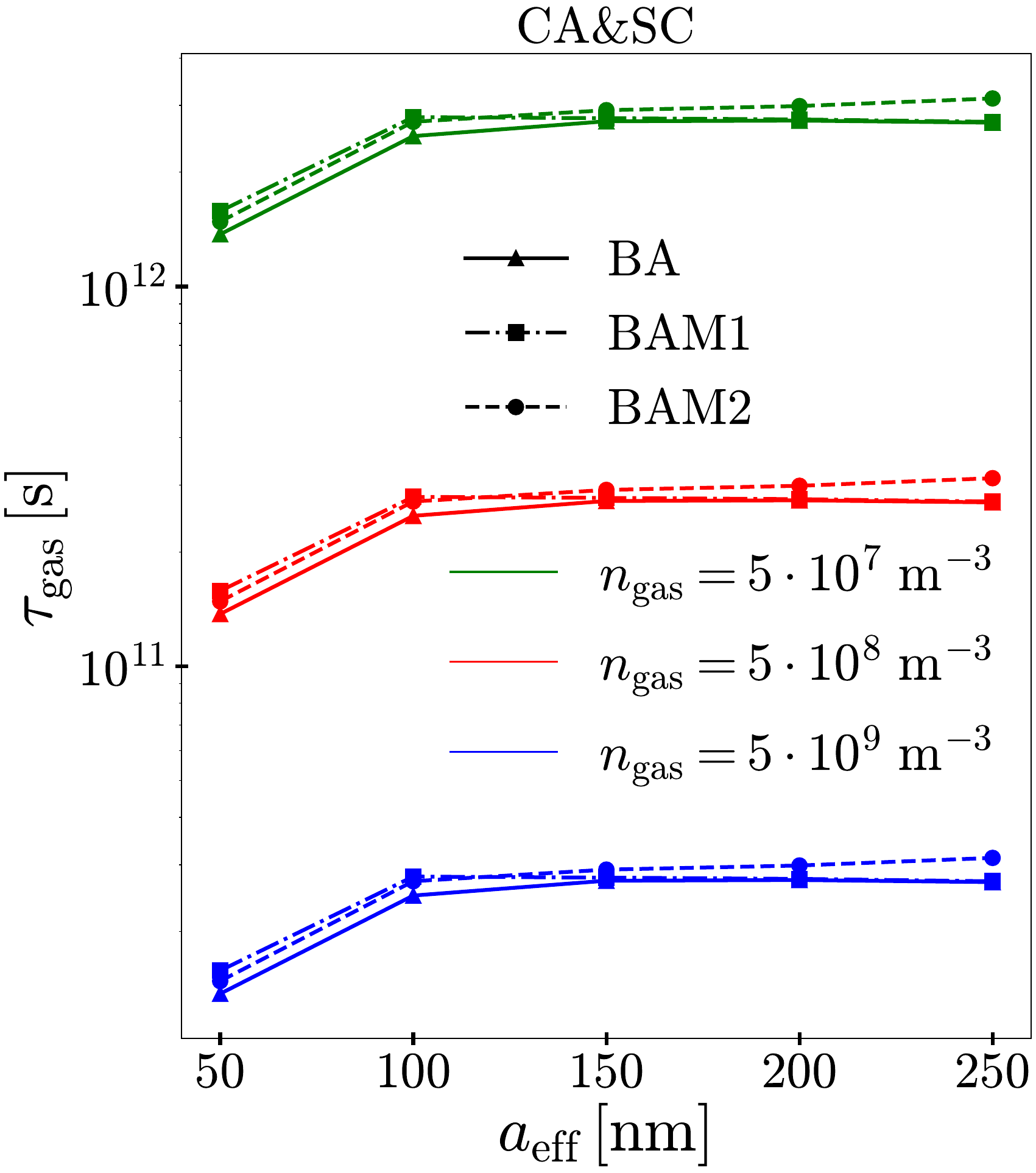}
    \includegraphics[width=0.33\textwidth]{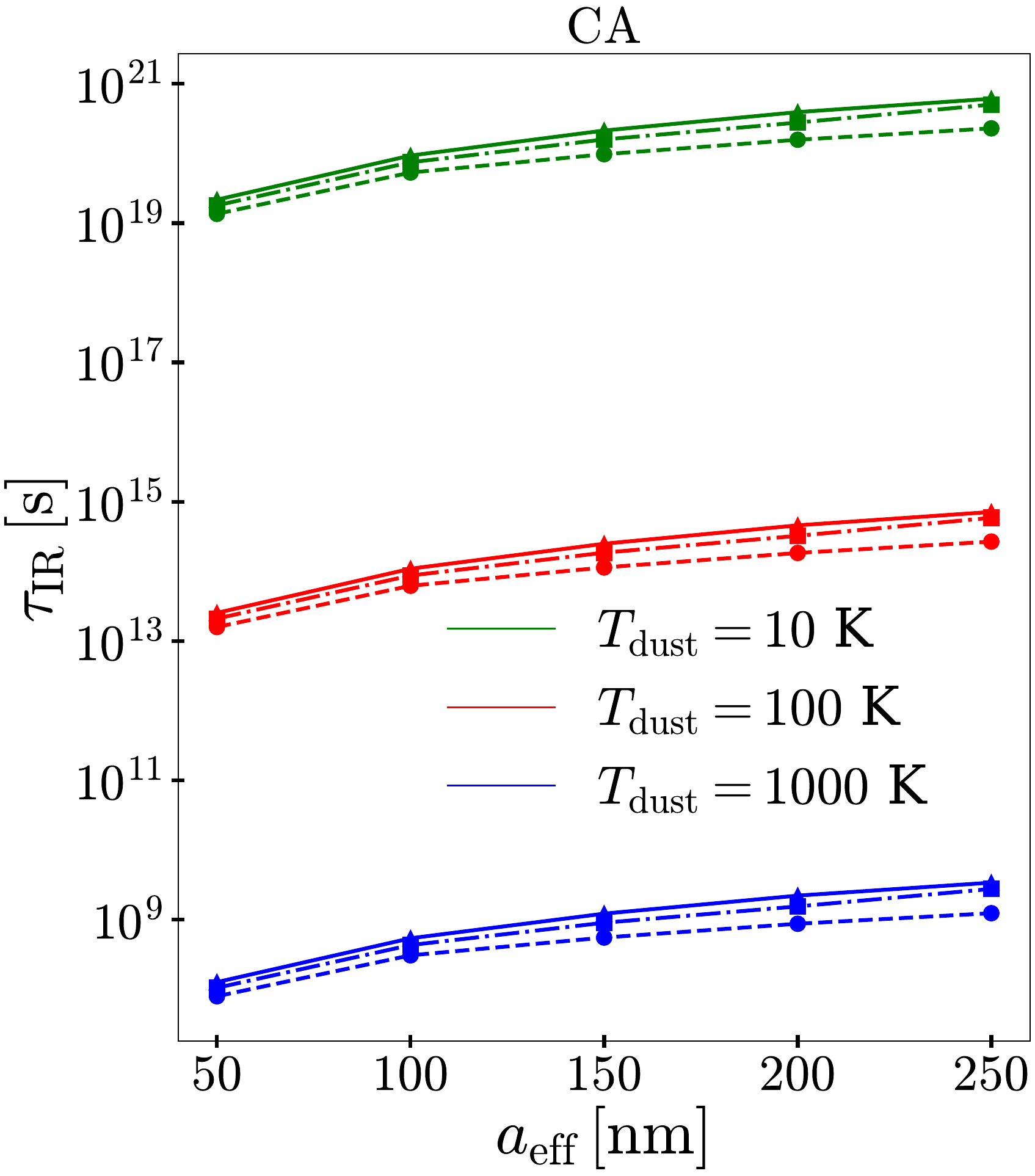}
    \includegraphics[width=0.33\textwidth]{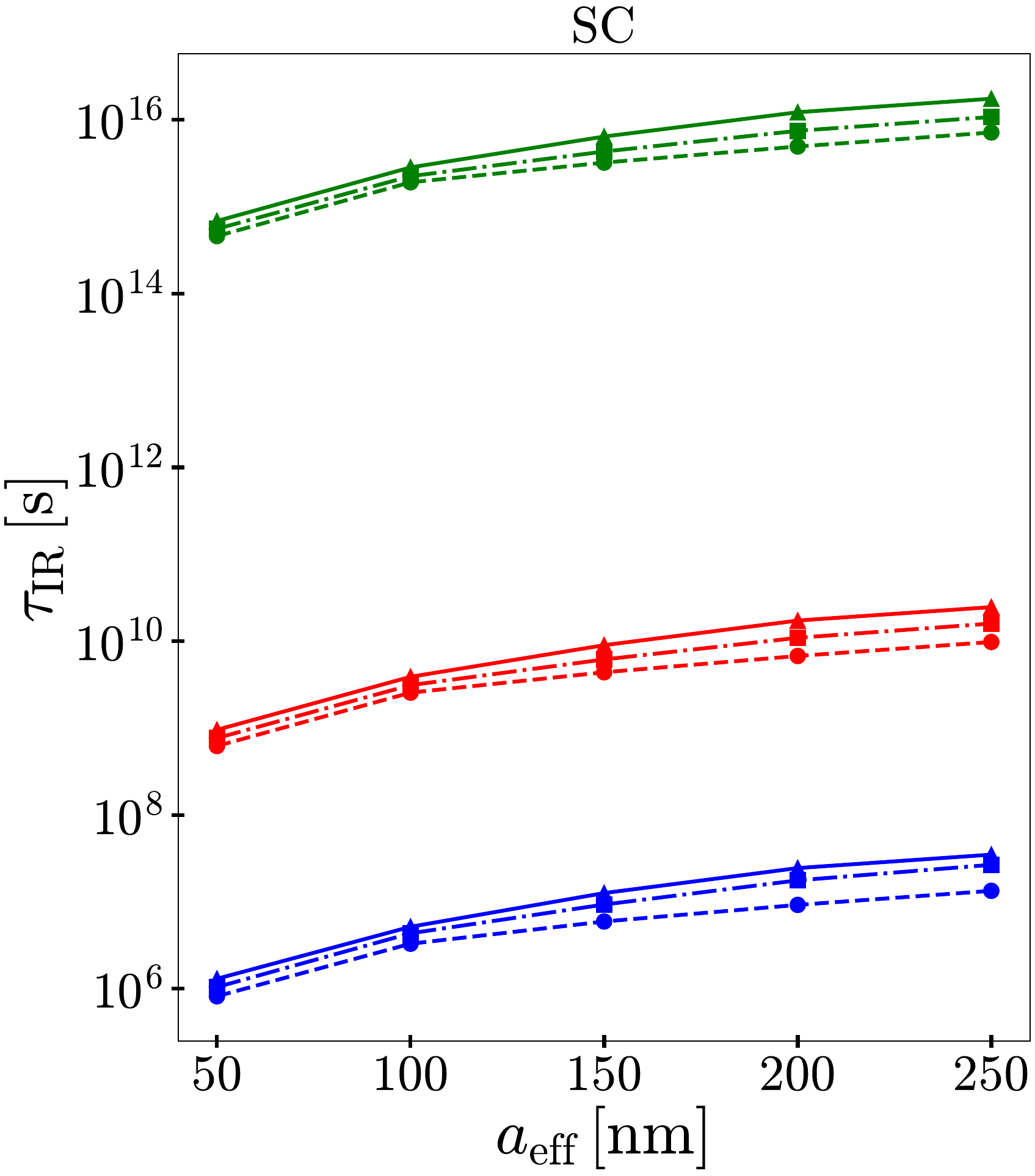}
	\caption{Drag timescales over the effective radius $a_{\mathrm{eff}}$ dependent on the grain model (BA, BAM1, or BAM2). \textit{Left panel:} Gas drag timescale for different gas densities $n_{\mathrm{gas}}$ for all grain materials. The gas temperature is at a  constant value of $T_{\mathrm{gas}}=100\ \mathrm{K}$.
    \textit{Center panel:} IR drag timescale for the CA grain material for different dust temperatures $ T_{\mathrm{dust}}$. \textit{Right panel:} Same as the center panel, but for the SC grain material.}
	\label{fig:drag_timescales}
\end{figure*}

\section{Results}\label{sect:results}
\subsection{Drag timescales}\label{subsect:drag_timescales}
We show the resulting timescales for the gas and photon drag for different gas densities and dust temperatures in Fig.~\ref{fig:drag_timescales}. The gas-drag timescale  $\tau_{\mathrm{gas}}$ is presented in the left panel as an average of SC and CA materials. We emphasize that the gas-drag efficiency $Q_{\mathrm{gas}}$ is only dependent on the grain shape in our modeling and is independent of the material. Denser gas equals a lower $\tau_{\mathrm{gas}}$, in agreement with the literature (see, e.g., \citealp{DraineWeingartner1997} and \citealp{Reissl2023}). The dependence on the monomer density is marginal, with the BA grains having a stronger gas drag than the BAM1 and BAM2 grains. This is due to the dependence of $Q_{\mathrm{gas}}$ on the surface area.\\
In detail, the BA grains present a larger surface area for interaction with gas particles. Following Eq.\ref{eq:tau_gas}, the expected size dependence of $\tau_{\mathrm{gas}}$ would be linear ($I_{\mathrm{max}}\propto a_{\mathrm{eff}}^5$). It is almost constant above $100\, \mathrm{nm}$, however, which is due to the linear increase in $Q_{\mathrm{gas}}$  with size. In principle, the gas drag efficiency $Q_{\mathrm{gas}}$ should only depend on the shape of the grain and not on the size.  $Q_{\mathrm{gas}}$ nonetheless increases with size, and the reason for this may be that $a_{\mathrm{eff}}$ is poorly suited to describe the grain size in this situation. The calculation of $Q_{\mathrm{gas}}$ used the simulated gas-drag-torque, which depends on the angular momentum lost by the grain when a gas particle leaves the surface. This in turn scales linearly with the distance from the axis of rotation of the grain from which the gas particle is ejected. For porous grains, this distance is often larger than $a_{\mathrm{eff}}$. The torque simulated in this way contains information about the true shape of the grain, but in the formula for $Q_{\mathrm{gas}}$ , the size $a_{\mathrm{eff}}$ is used, which creates the size dependence of $Q_{\mathrm{gas}}$.\\

The photon-drag timescale is shown in the middle for CA grains and on the right for SC grains in Fig.~\ref{fig:drag_timescales}. As the efficiency of absorption $Q_{\mathrm{abs}}$ is higher overall for SC, more photons are absorbed and emitted, which causes a stronger photon drag. The number of emitted photons also depends on the dust temperature, with hotter dust experiencing stronger drag (see also Tab.~\ref{tab:T_dust}). The dependence on the monomer density is marginal, and this is related to the fact that the moment of inertia is higher in the case of BA grains than for the more compact BAM2 grains for the same effective radius \citep[see Appendix A in][]{Reissl2023}.

\begin{figure}[htbp]
    \centering
    \includegraphics[width=0.48\textwidth,left]{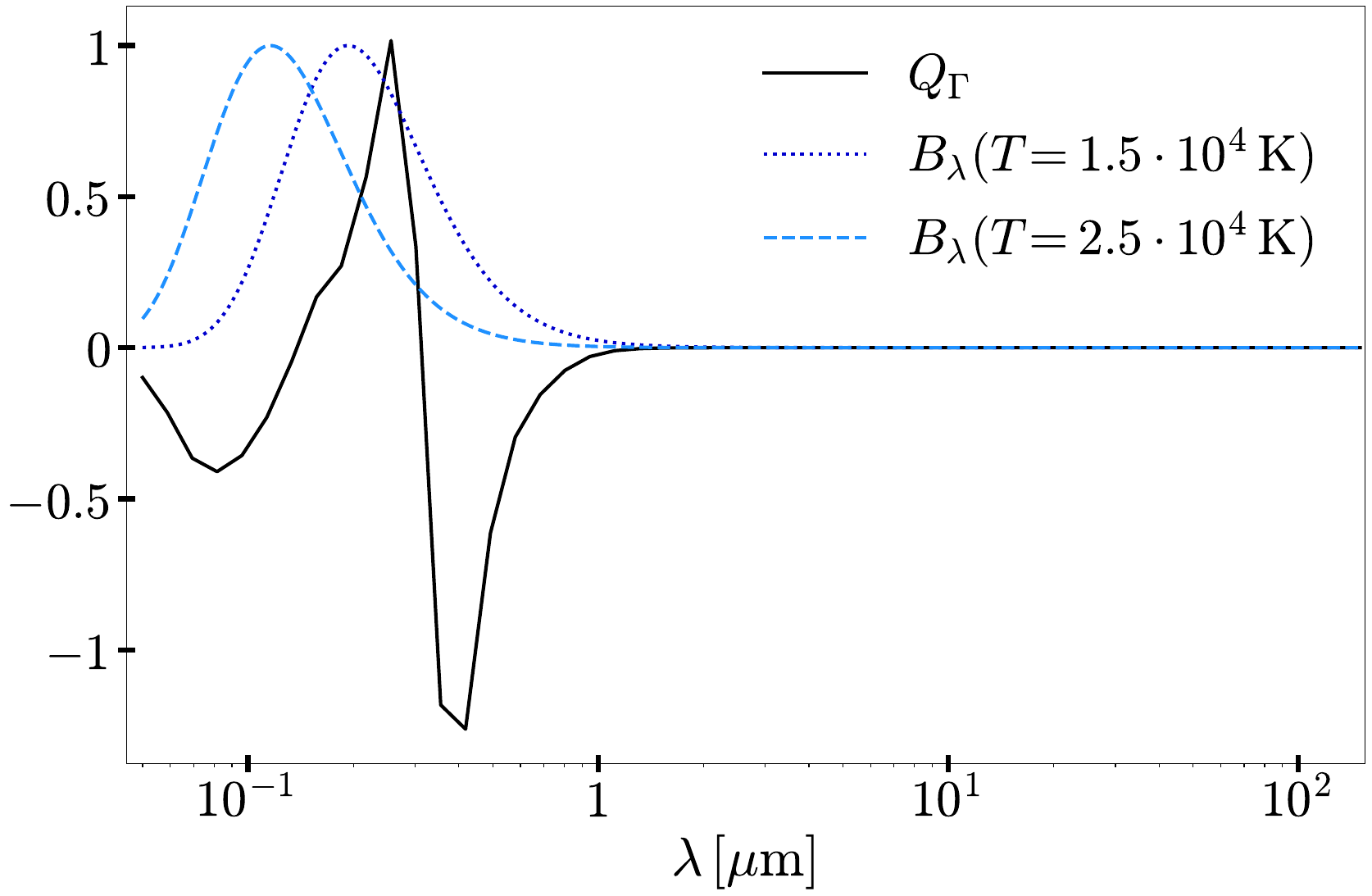}
	\caption{Torque efficiency $Q_\Gamma$ as a function of wavelength $\lambda$ for an exemplary CA BA grain of size $a_{\mathrm{eff}}=200\,\mathrm{nm}$. The spectra of the two stellar sources S1 and S2 are shown in blue, normalized by their maximum value. We emphasize that this particular grain has two changes in sign, resulting in a reduced net torque over wavelengths.}
    \label{fig:example_grain}
\end{figure}

\begin{figure*}[htbp]
	\centering
    \includegraphics[width=0.49\textwidth]{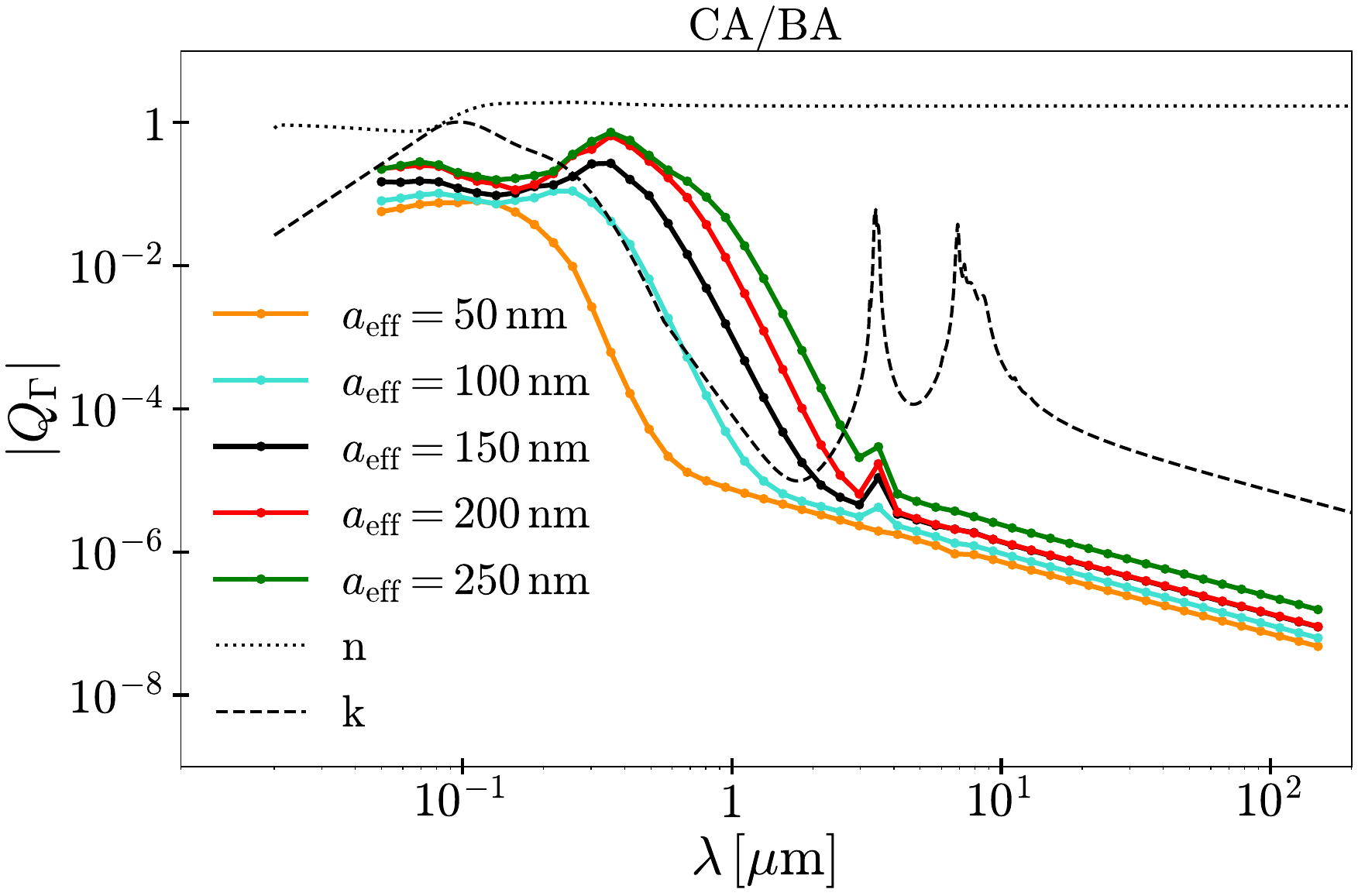}
	\includegraphics[width=0.49\textwidth]{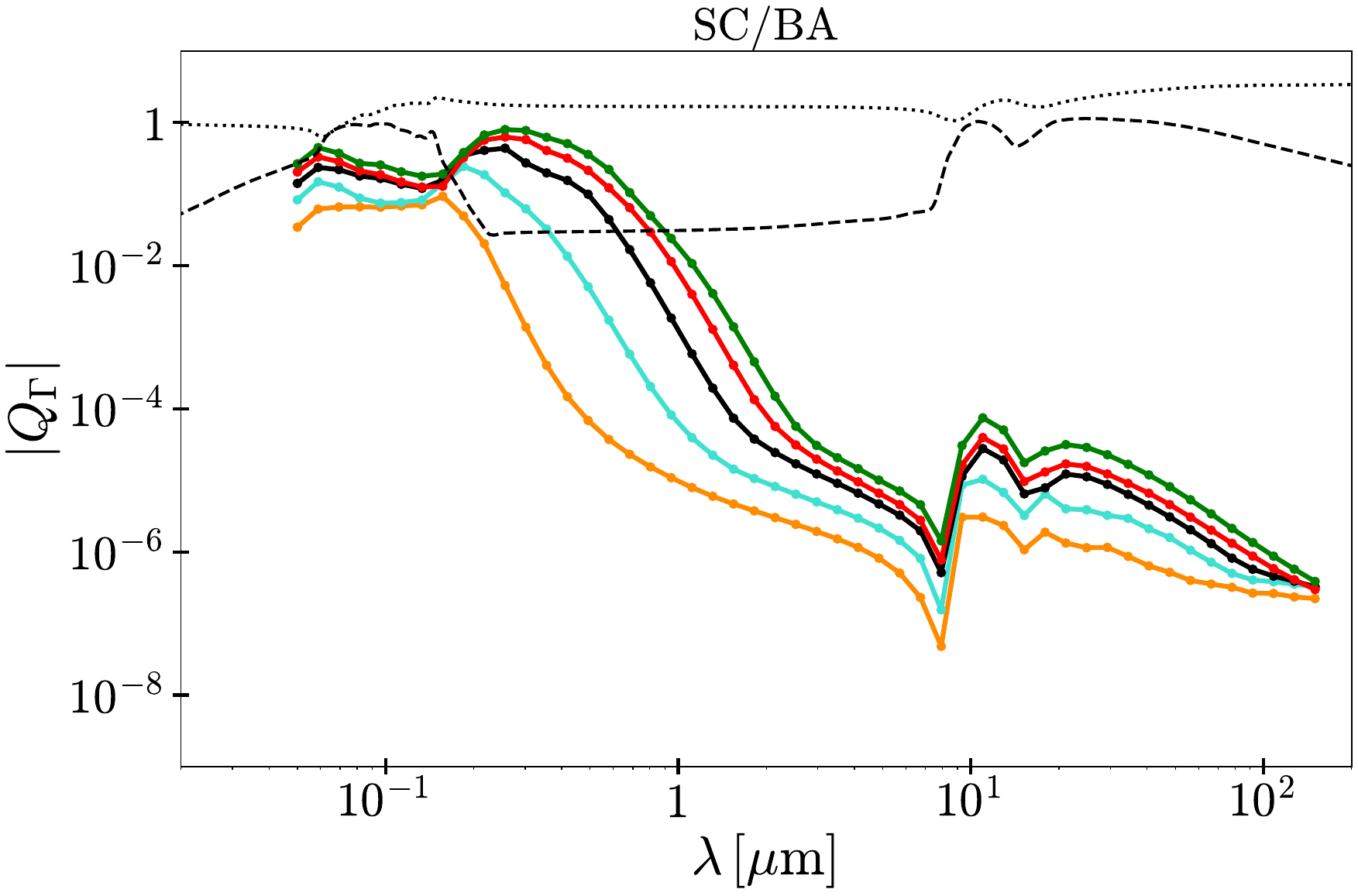}
	\caption{Average torque efficiency $|Q_\Gamma|$ over wavelengths for CA (left) and SC (right) BA aggregates of different grain sizes $a_{\mathrm{eff}}$ (color-coded) in comparison with the real part $n$ (dotted line) and imaginary part $k$ (dashed line) of the corresponding refractive indices. We note that the characteristic features of $Q_\Gamma$ follow the shape of $k$.}
	\label{fig:Qav_nk}
\end{figure*}

\begin{figure*}[htbp]
	\centering
	\includegraphics[width=0.49\textwidth]{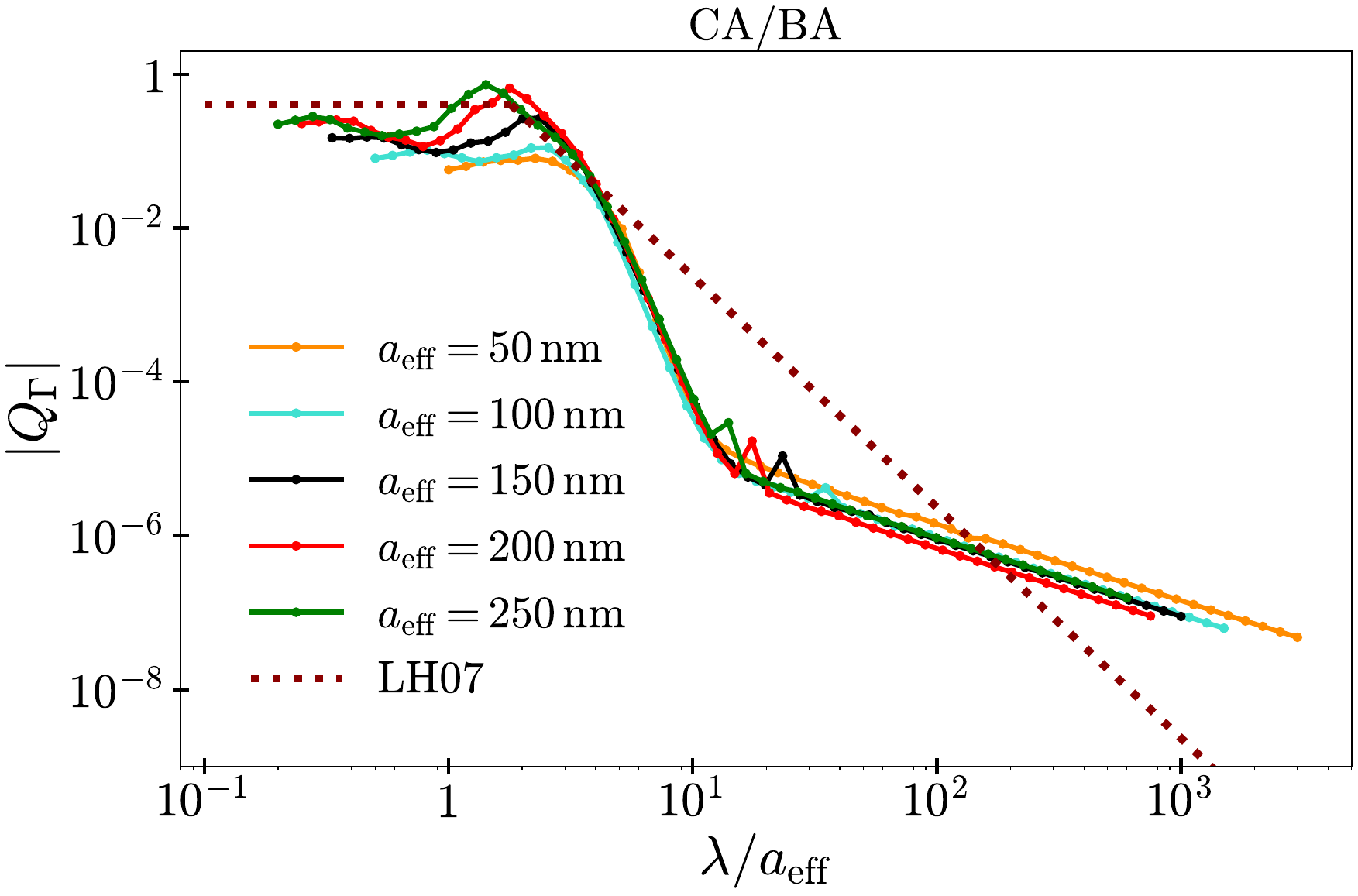}
    \includegraphics[width=0.49\textwidth]{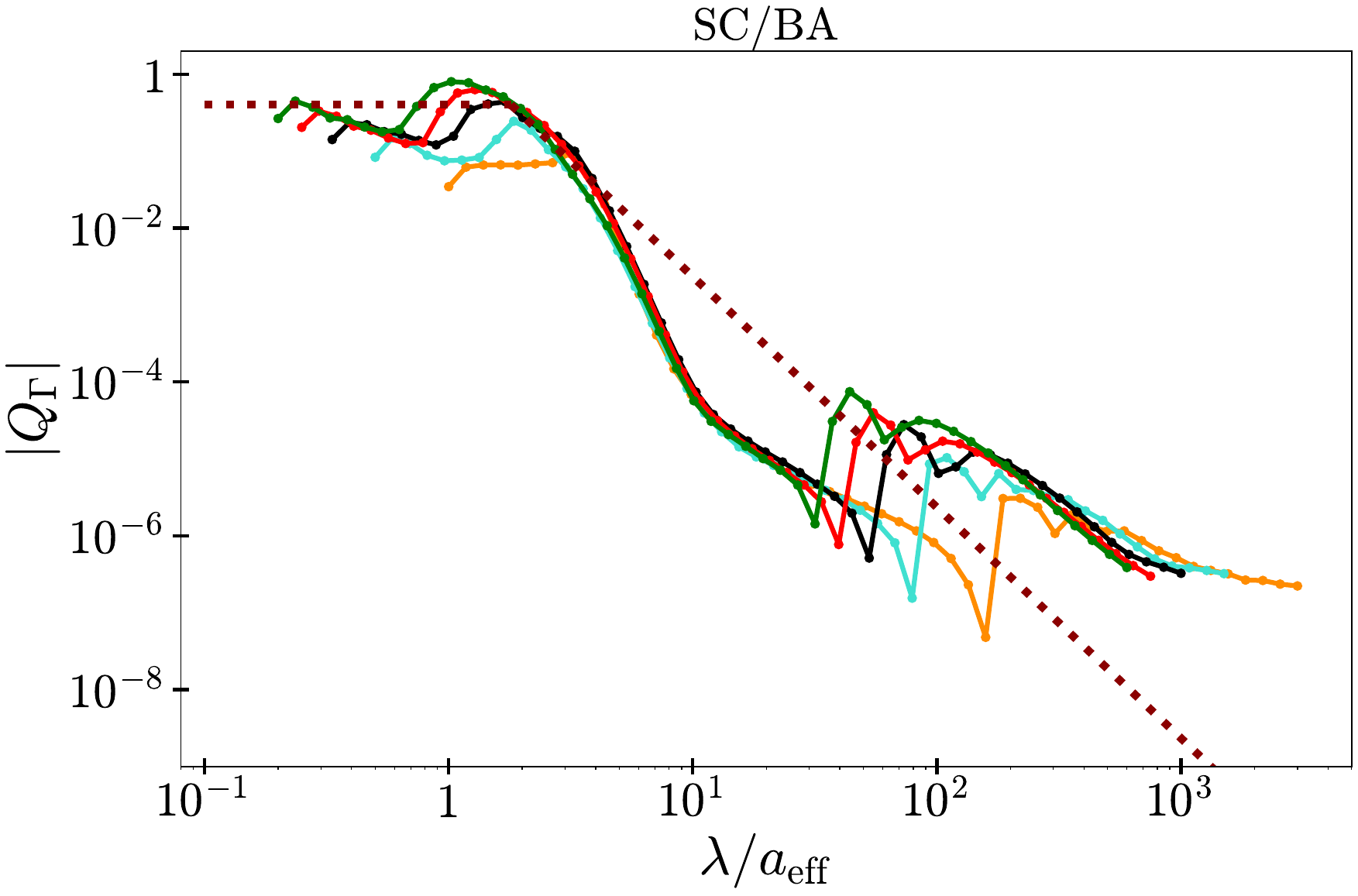}
	\caption{Same as Fig.~\ref{fig:Qav_nk}, but for the average torque efficiency $Q_\Gamma$ over $\lambda/a_{\mathrm{eff}}$ in comparison  with the conical power-law parameterization (dotted red line), as suggested in LH07.}
	\label{fig:Qav_can}
\end{figure*}

\begin{figure*}[htbp]
	\centering
	\includegraphics[width=0.49\textwidth]{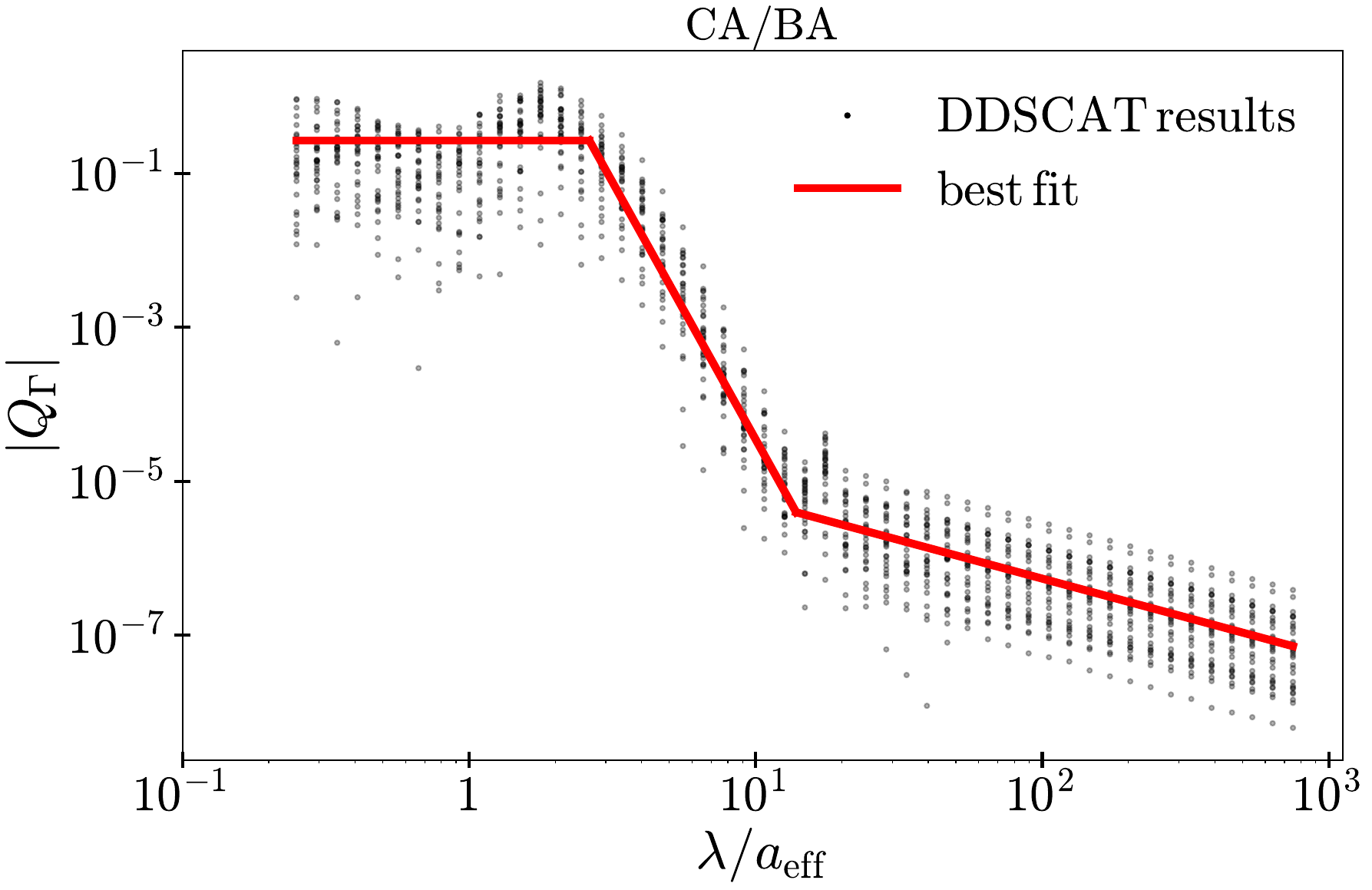}
    \includegraphics[width=0.49\textwidth]{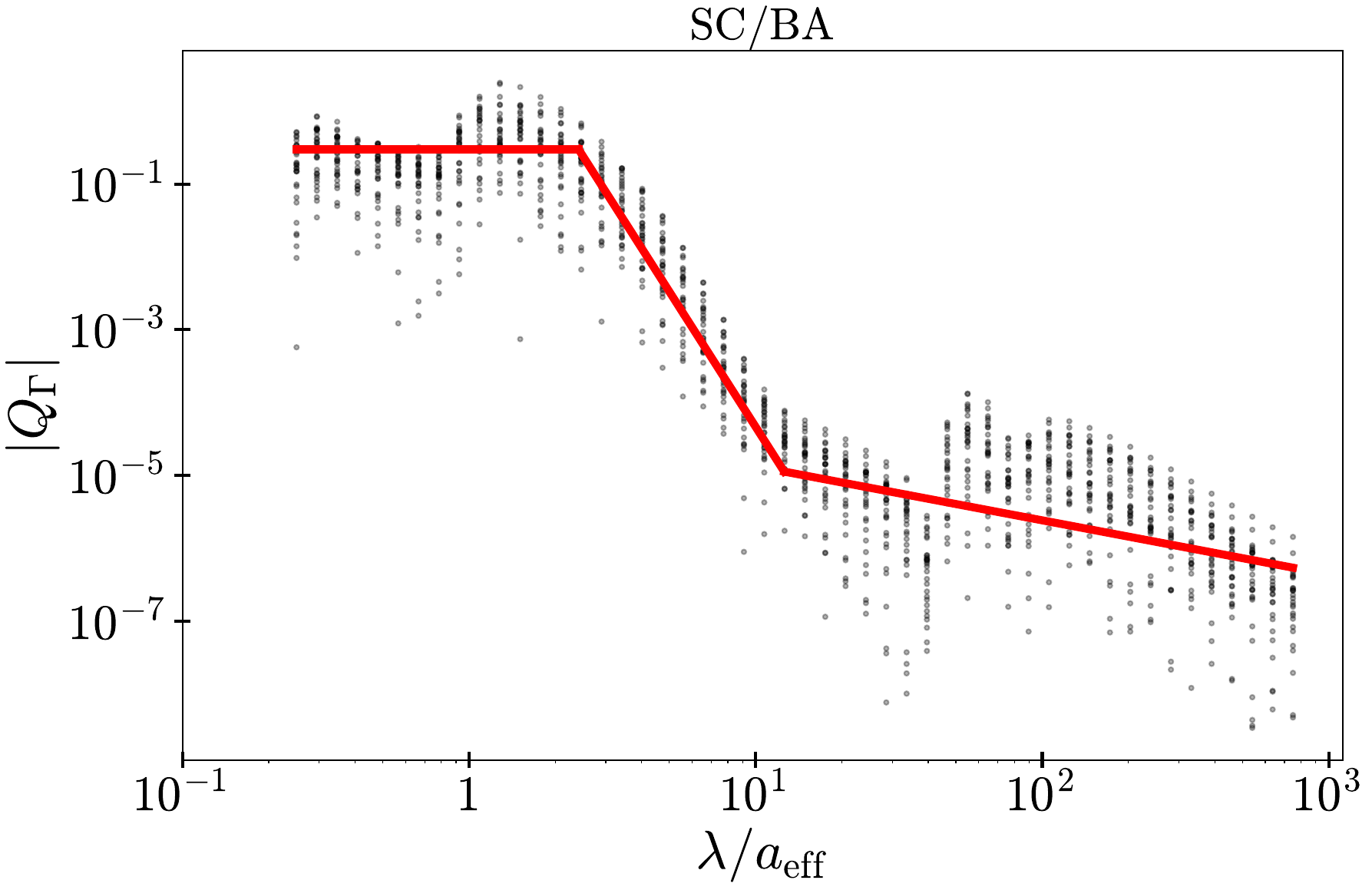}
	\caption{Two-power-law parameterization shown as a red line for CA (left) and SC (right) BA grains in comparison with the full set of the numerical DDSCAT simulations (black dots) for grains of size $a_{\mathrm{eff}}=200\ \mathrm{nm}$, where the absolute value of the torque efficiency $|Q_\Gamma|$ is plotted as a function of $\lambda/a_{\mathrm{eff}}$. The constant part $c$ is fitted for each grain size up to a value of $\lambda/a_{\mathrm{eff}}=3$, the steeper power-law is fit as a linear function to the logarithm of the data for all sizes combined in the interval $3<\lambda/a_{\mathrm{eff}}<10$, and the same for the shallow power law in the interval $\lambda/a_{\mathrm{eff}}>20$. Finally, the two intersection points $a$ and $b$ were calculated and used to create a continuous parameterization for $|Q_\Gamma|$ (see Eq.~\ref{eq:2pp}).}
	\label{fig:Q_2pp}
\end{figure*}


\subsection{Torque efficiency}
As we assumed that the principal grain axis $\hat{a}_1$ is parallel to $\hat{e}_1$, the direction of the incoming radiation, only the $Q_1$ component needs to be evaluated to obtain the total torque efficiency $Q_\Gamma$.\\
The results of the efficiency $Q_\Gamma$ calculated with DDSCAT may possess positive (negative) values depending on the wavelength (see Fig.~\ref{fig:example_grain}), corresponding to a direction of the angular velocity $\vec{\omega}_{\mathrm{RAT}}$ that is parallel (antiparallel) to $\hat{e}_1$. Hence, different wavelength regimes create opposing torques on the grain aggregates. When the sign change is subsequently taken into account, the net RAT for individual aggregates may considerably lower. In order to evaluate the average efficiency for different grain aggregates, we evaluated $\Gamma_{\mathrm{RAT}}$ over $\lambda$ (see Eq.~\ref{eq:GammaRAT}) for each grain individually, including canceling effects of positive and negative spectral contribution of $Q_\Gamma(\lambda)$, and we also 
calculated $\Gamma_{\mathrm{RAT}}$ using the absolute value $|Q_\Gamma(\lambda)|$, which ignores the opposing spectral contributions to the net RAT. We find that averaging over the DDSCAT data with the absolute value of $|Q_\Gamma|$ resulted in only marginally higher RATs compared to when the sign of $Q_\Gamma$ was taken into account. We emphasize that the difference is insignificant compared to the variation between individual grain shapes. Hence, we only used the absolute value $|Q_\Gamma|$ in the following analysis so that data of different grain shapes can be combined.

The resulting dependence of the magnitude of $Q_\Gamma$ on the wavelength for CA BA and SC BA grains is presented in Fig.~\ref{fig:Qav_nk}.  The values of $Q_\Gamma$  are compared with the refractive index of the corresponding grain material. We report a feature for SC at $\lambda > 8\,\mu\mathrm{m}$ that follows the form of the imaginary part $k$ of the refractive index. This feature of SC was also noted in previous studies (see, e.g., LH07).

\subsection{Parameterization for the RAT efficiency}
Since RATs depend on the wavelength $\lambda$ relative to the grain radius $a_{\mathrm{eff}}$, it is convenient to evaluate $|Q_\Gamma|$ as a function of $\lambda/a_{\mathrm{eff}}$ (see Fig.~\ref{fig:Qav_can}) instead of $\lambda$ alone. Here, the slopes of the different grain sizes overlap tightly, showing the self-similar nature that allowed the canonical parameterization (Eq.~\ref{eq:QCanonical}) for solid grains in the first place. Our results using grain aggregates show a comparable constant part for $\lambda/a_{\mathrm{eff}}<1.8$, but do not follow the overall power-law parameterization for $\lambda/a_{\mathrm{eff}}>1.8$. Instead, both materials SC and CA reveal a steeper power law that overlaps for all sizes from $\lambda/a_{\mathrm{eff}}\approx 3$ and then evolves into a shallower power law for $\lambda/a_{\mathrm{eff}}
\gtrsim 10$. A similar behavior of a broken power law was reported in previous studies, for instance, in LH07, or for Gaussian dust grains in \cite{herranen_radiative_2019}, but is much more apparent here. Consequently, we suggest a modified parameterization for ballistic grain aggregates with two connected consecutive power laws. We performed least-square fits for the constant part (called $c$ in this study) and the two power laws for all data points of the grains that shared the same material and monomer density. The fits of $c$ were performed for individual grain sizes in the interval $\lambda/a_{\mathrm{eff}}<3$. All sizes were fit jointly for the first and second power law with exponents $q_1$ and $q_2$ in the intervals $3<(\lambda/a_{\mathrm{eff}})<12$ and $(\lambda/a_{\mathrm{eff}})>20$. In practice, the power laws were fit as linear functions to the logarithm of the data.

In order to conveniently combine the three regions into a continuous curve, we chose the following definition:
\begin{equation}
|Q_\Gamma|\;=
 \begin{cases} 
    c\left(a_{\mathrm{eff}}\right) &\mbox{if } \left(\frac{\lambda}{a_{\mathrm{eff}}}\right) \leq a \, ,\\ 
        c\left(\frac{\lambda}{a_{\mathrm{eff}}}/a\right)^{q_1}  &\mbox{if } a \leq \left(\frac{\lambda}{a_{\mathrm{eff}}}\right) \leq b\, , \\ 
		c\left(\frac{b}{a}\right)^{q_1}\left(\frac{\lambda}{a_{\mathrm{eff}}} /b\right)^{q_2}  &\mbox{otherwise}\, .
    \end{cases}
\label{eq:2pp}
\end{equation}
We joined the individual fits by calculating the intersection points $a$ and $b$, where $a$ is the intersection of $c$ with the steep power law, and $b$ is the intersection of both power laws. Therefore, like $c$, there will be a different value of $a$ for each $a_{\mathrm{eff}}$. The resulting parameterization is shown in Fig.~\ref{fig:Q_2pp} for the exemplary data set of carbon BA grains of size $200\, \mathrm{nm}$. All resulting fit parameters of Eq.~\ref{eq:2pp} are given in Appendix \ref{ap:2pp_params}.

Larger grains interact more efficiently with the radiation field, which results in a higher value $c$, as shown in Fig.~\ref{fig:Qav_can}. The BAM1 and BAM2 grain models have comparable curves for $|Q_\Gamma|$, where the values differ most in the constant part on a scale that is similar to the difference between two grain sizes. The more porous the grain, the larger $c$. The two materials differ in their values of $c$, which are higher for SC. Both power laws are steeper for CA, most likely due to interference by the feature in the SC material.

\subsection{Net torques and angular velocities}
The exact difference in RAT using the canonical or the modified parameterization, and subsequently, the magnitude of the angular velocity of the grains, cannot a priori be quantified because the RATs depend on the spectral characteristics of the radiation field. Therefore, we calculated the maximum angular velocity $\omega_{\mathrm{RAT}}$ induced by RATs using both parameterizations and compared the results to those we obtained using the DDSCAT data of individual grains.

If the torque efficiency $Q_\Gamma(\lambda)$ for a single grain exhibits some sign changes over wavelength (see Fig.~\ref{fig:example_grain}), this leads to a lower torque overall as different wavelengths exert RAT components in opposite directions. As mentioned above, the effect of positive and negative components of $Q_\Gamma$ canceling out appears to be small compared to the overall variation in the strength of $Q_\Gamma$ between different grains. 
The torque along $\hat{a}_1$ was calculated for each grain following Eq.~\ref{eq:GammaRAT}, where $Q_\Gamma$ was used without taking the absolute value. Then, the maximum angular velocity was calculated with Eq.~\ref{eq:omegaRAT}, where $\tau_{\mathrm{drag}}$ and $I_{\mathrm{max}}$ were also evaluated individually for each grain.

\begin{figure*}[htbp]
	\centering
    \includegraphics[width=0.49\textwidth]{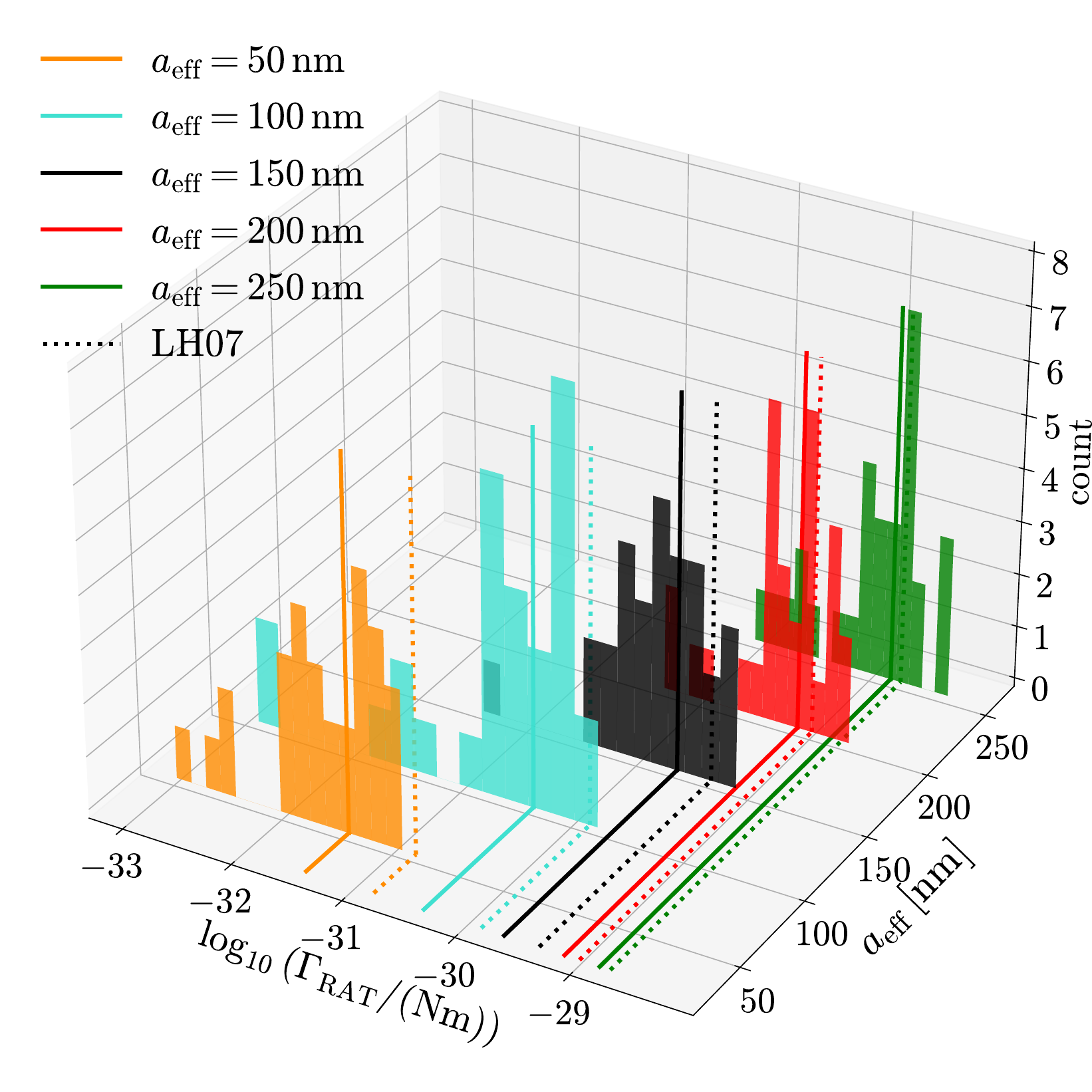}
    \includegraphics[width=0.49\textwidth]{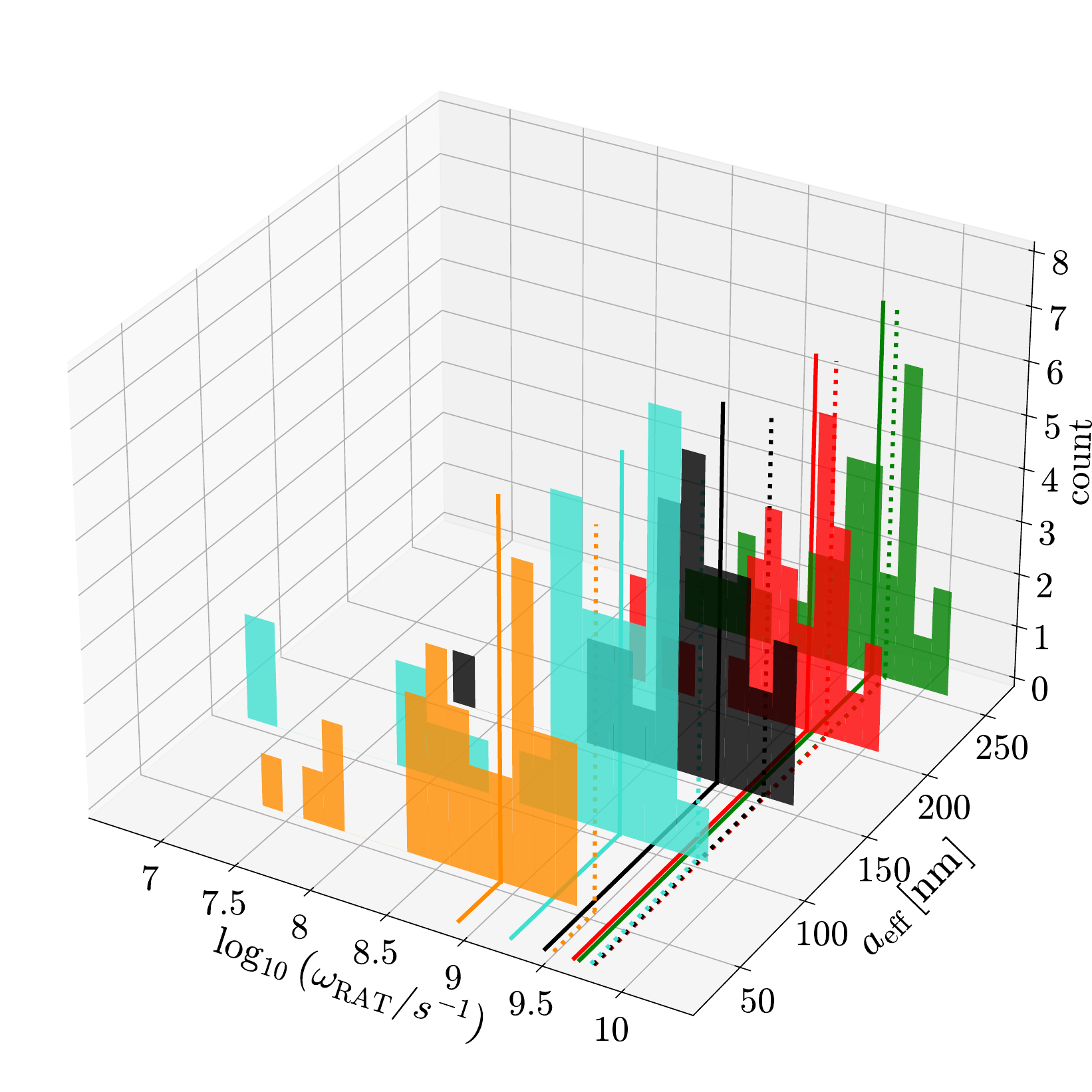}
    \caption{Histograms representing the distribution of the RAT $\Gamma_{\mathrm{RAT}}$ (left) and angular velocity $\omega_{\mathrm{RAT}}$ (right) for SC BA grains of different sizes and the AGN as the radiation source. The continuous lines show the mean for each distribution, and the dotted lines show the value calculated using the canonical parameterization (LH07). The mean values for $\omega_{\mathrm{RAT}}$ are calculated from the mean values of $\Gamma_{\mathrm{RAT}}$ shown in the lower panel as well as averages for $\tau_{\mathrm{drag}}$ and $I_{\mathrm{max}}$. We note that the RATs by the canonical parameterization are consistently higher than the mean values of the dust aggregates.}
    \label{fig:Hists}
\end{figure*}

Figure \ref{fig:Hists} shows the distribution of  angular velocities and torques for individual SC BA grains and the most luminous source, that is, the AGN. The distributions are broken down by grain size. We report that the resulting values of the RATs may span a range of up to two orders of magnitude, with a larger spread below the mean. The mean of the distribution increases with grain size because larger grains have higher $Q_\Gamma$ values in the constant part. The distributions of the different grain sizes are more distinct in $\Gamma_{\mathrm{RAT}}$ and overlap significantly for $\omega_{\mathrm{RAT}}$ because of the drag timescale (see Eq.~\ref{eq:omegaRAT}).

A parameterization that describes the average value for an ensemble of grains means that we do not need to calculate $\omega_{\mathrm{RAT}}$ for every single grain. We therefore compared the average values obtained from individual grain calculations (shown in Fig.~\ref{fig:Hists} as solid lines) with the values obtained for the canonical parameterization for $Q_\Gamma$ (dotted lines). This parameterization does not model the means of the distributions well. It overestimate the values of $\Gamma_{\mathrm{RAT}}$ and $\omega_{\mathrm{RAT}}$, especially for the smaller grains.

To quantify how the canonical parameterization of LH07 and the modified parameterization of our study compare with the distribution means for all combinations of radiation sources, materials, and monomer densities, we show the distribution means of the RATs in Fig.~\ref{fig:av_trq+2p_para} as a function the grain size $a_{\mathrm{eff}}$. The CA material in the left column, the SC material in the right column, the monomer densities in the rows, and the four radiation sources are shown with different colors in each panel. The difference between the sources is due to the linear dependence of $\Gamma_{\mathrm{RAT}}$ on the energy density $u$ and anisotropy factor $\gamma$,
where larger RATs are reached for stronger radiation sources (see Eq.~\ref{eq:GammaRAT} and Tab.~\ref{tab:sources}). The RAT of the S1 source should be larger by a factor of 7 due to $\gamma$ and by a factor of $5.6$ due to $u$. The difference seen in Fig.~\ref{fig:av_trq+2p_para} is larger than that with approximately two orders of magnitude. This is a result of the spectral energy distribution of this source. The integral in Eq.~\ref{eq:GammaRAT} places a much larger weight on $u$ at wavelengths shorter than $\approx 1\,\mu\mathrm{m}$, as $Q_\lambda$ drops by about four orders of magnitude for longer wavelengths. In this range, the stellar sources outshine the ISRF.

In Fig.~\ref{fig:av_wrat+2p_para} we present the mean angular velocities $\omega_{\mathrm{RAT}}$ corresponding to the RATs shown in Fig.~\ref{fig:av_trq+2p_para}.  Here, the variations in $\omega_{\mathrm{RAT}}$ between individual sources are different from the variation in $\Gamma_{\mathrm{RAT}}$ because $\omega_{\mathrm{RAT}}$ also depends on $\tau_{\mathrm{drag}}$, which itself depends indirectly on $u$ via the temperature dependence of $\tau_{\mathrm{IR}}$. Using the dust temperatures in Tab.~\ref{tab:T_dust}, we can estimate $\tau_{\mathrm{IR}}$ for the given temperature from Fig.~\ref{fig:drag_timescales}. The gas-drag timescale is independent of $u$ and about $\tau_{\mathrm{gas}}\gtrsim 10^{12}\,\mathrm{s}$ for the chosen gas density of $n_{\mathrm{gas}}=5\cdot 10^7\,\mathrm{m^{-3}}$. The drag timescale is dominated by $\tau_{\mathrm{gas}}$ for the lower dust temperatures of the ISRF and S1 sources, both timescales are about equal for S2, and $\tau_{\mathrm{IR}}$ dominates for the AGN, where the dust is hottest.

Therefore, the difference in $\omega_{\mathrm{RAT}}$ only depends on $\Gamma_{\mathrm{RAT}}$ for the ISRF and S1, but it includes $\tau_{\mathrm{IR}}$ for S2 and AGN. The AGN therefore is much closer to S2 in $\omega_{\mathrm{RAT}}$ ($1\,\mathrm{dex}$) than in $\Gamma_{\mathrm{RAT}}$ ($3\,\mathrm{dex}$). Even though the torque is much stronger for the AGN source, the grains do not rotate that much faster because they also become much hotter, which increases the IR drag. The difference in $\tau_{\mathrm{IR}}$ between the two materials also leads to a lower angular velocity for S2 and AGN for SC compared to CA, but there is no difference between the two materials for S1 and ISRF because $\tau_{\mathrm{gas}}$ dominates there. The difference between monomer densities is a decrease in $\Gamma_{\mathrm{RAT}}$ from BA to BAM2 that is smaller than one order of magnitude because $Q_\Gamma$ decreases with monomer density. The resulting decrease with monomer density in $\omega_{\mathrm{RAT}}$ is smaller because $I_{\mathrm{max}}$ also decreases with monomer density.

As expected, RATs strongly scale with $a_{\mathrm{eff}}$, and this does not change much between sources, materials, and monomer densities. However, for the angular velocity $\omega_{\mathrm{RAT}}$, the results are less uniform, as shown in Fig.~\ref{fig:av_wrat+2p_para}. Here, there is no scaling with grain size for the ISRF source and a weak scaling for the AGN, which flattens toward the larger grains. For the stellar sources, $\omega_{\mathrm{RAT}}$ first increases and then flattens for SC, and it even decreases toward larger grains for S1. In the case of CA, $\omega_{\mathrm{RAT}}$ even scales inversely with grain size for the stellar sources.
We remark that the scaling of $\omega_{\mathrm{RAT}}$ with grain size depends on the interplay between multiple factors: The spectral shape of the source and the scaling of the drag timescale, which depends on the material and structure of the grain, and whether $\tau_{\mathrm{IR}}$ or $\tau_{\mathrm{gas}}$ dominates for the given parameters.
As outlined in the pioneering work of \citet{Hoang2019} and \citet{Hoang2019ApJ}, RATs may spin up dust grains to a characteristic angular velocity $\omega_{\mathrm{disr}}$, where the internal grain structure can no longer resist the disruption because of centrifugal forces (RATD). The predicted order of magnitude 
for RATD to occur is $\omega_{\mathrm{disr}}\approx 10^9 \mathrm{s}^{-1}$. Later, ab initio numerical N-body simulations of the disruption of rotating BA aggregates revealed that aggregates larger than $a_{\mathrm{eff}}=300\ \mathrm{nm}$ are more resistant against RATD because they are deformed. We emphasize that the most luminous sources, AGN and S2  may spin up the dust aggregates beyond the limit of $\omega_{\mathrm{RAT}}>\omega_{\mathrm{disr}}$ (see Fig.~\ref{fig:av_wrat+2p_para}), even though RATs are less efficient for ballistic aggregates. Hence, most of the grains may become disrupted by RATD in the proximity of an AGN or S2-like star. This means that ballistic aggregates are more likely to grow deep inside of dense molecular clouds, where they are shielded from a strong radiation field \citep[][]{Bethell2007ApJ}.\\
As presented in Fig.~\ref{fig:av_wrat+2p_para}, the canonical parameterization gives more accurate results for the larger grain aggregates because they experience the strongest torques. We point out that the canonical parameterization overestimates the mean angular velocity $\omega_{\mathrm{RAT}}$ overall, with differences of up to one order of magnitude. In comparison, the two-power-law parameterization resembles the mean of the DDSCAT data more precisely and performs equally well for all sources, materials, monomer densities, and sizes.

A way to quantify the efficiency of RATs of porous aggregates was proposed by \cite{tatsuuma_rotational_2021} (TK21 hereafter). The TK21 parameterization takes the porosity (monomer density) of the grain aggregates into account via the volume filling factor $\phi$, where $\phi=1$ represents a solid grain, and $\phi=0$ represents completely empty space. Here, the volume-filling factor $\phi$ was added to the canonical parameterization as follows:
\begin{equation}
    |\vec{Q}_\Gamma| =
    \begin{cases} 0.4  &\mbox{if } \lambda \leq 1.8\,a_{\mathrm{eff}}\cdot \phi \, ,\\ 
	0.4 \left(\frac{\lambda}{1.8\, a_{\mathrm{eff}}\,\cdot\,\phi  } \right)^{\alpha}  & \mbox{otherwise}\, .\end{cases}
 \label{eq:QTatsuuma}
\end{equation}
Fig.~\ref{fig:av_wrat+Tatsuuma_para} shows the average angular velocity $\omega_{\mathrm{RAT}}$ in comparison with the TK21 parameterization and canonical parameterization of LH07. The model proposed by TK21 clearly improves upon the canonical parameterization of LH07 for the less porous BAM1 and BAM2 grains, but now slightly underpredicts $\omega_{\mathrm{RAT}}$ for the weaker sources in BAM1. For the more porous BA grains, however, TK21 clearly underpredicts $\omega_{\mathrm{RAT}}$ because in this regime, the canonical parameterization performed best. As shown in Fig.~\ref{fig:av_wrat+Tatsuuma_para}, merely introducing $\phi$ to take the grain porosity into account does not bring the canonical parameterization into agreement with our results. This indicates that complex grain shapes differ in their interaction with radiation compared to compact grains in more fundamental ways than just by making the grains porous.

Consequently, we may summarize that even though there is a large spread in $\omega_{\mathrm{RAT}}$ between individual grains (see Fig.~\ref{fig:Hists}) due to random variation in the grain shape, it is still valid to use a parameterization of $Q_\Gamma$ to model the average $\omega_{\mathrm{RAT}}$ for an ensemble of ballistic grain aggregates.\\ For the three parameterizations shown in this work, the two-power-law parameterization is the most accurate choice for this task.
 
\section{Discussion and caveats} \label{sect:discussion}
An underlying assumption in our RAT calculations using grain aggregates is the anisotropy of incoming radiation. As DDSCAT only simulates anisotropic radiation, we used the anisotropy factor $\gamma$ to treat a certain percentage of the source radiation as anisotropic. For instance, a grain in a molecular cloud around a forming star receives a portion $\gamma$ of the stellar radiation directly (anisotropically) and the rest isotropically from scattering on surrounding particles. These processes are not fully accounted for in this work, but we do not expect torques from isotropic radiation to dominate torques from anisotropic radiation, especially for our grain model without variations in the optical properties across the grain's surface.

We also assumed perfect alignment of the grain rotation axis with the direction of the radiation, that is, $\Theta=0$. This is based on the fact that grains preferentially align with the direction of incoming radiation in the absence of magnetic fields (see LH07), even though  magnetic fields are present in the typical environments probed here and the grains may not reach suprathermal angular velocities. For a more detailed picture, DDSCAT simulations that sample the whole range of $\Theta$ angles are required that are subsequently used to calculate phase-space trajectories of grains in $\Theta$ and $\omega$, and they would also need to include typical magnetic fields. This approach may allow us to verify whether the assumption of a fixed orientation was justified, and if not, to gauge how large the deviation is. However, this is beyond the scope of our current study.

It is still possible to use a parameterization to model the average angular velocity with the more detailed grain model, even including positive and negative values of $Q_\Gamma(\lambda)$ for individual grains. If needed, it might also be possible to parameterize the distribution of $\omega_{\mathrm{RAT}}$ values (see Fig.~\ref{fig:Hists}), or other quantities using larger grain ensembles.

As we were limited to a small range of grain sizes due to the computational restrictions of DDSCAT, future simulation setups should be designed to test whether the two-power-law parameterization for $|Q_\Gamma|$ also holds for larger grains. If this is the case and the two-power-law shape is also seen for general orientations of the grain, it would stand to reason that the form of $|Q_\Gamma|$ following two power laws is intrinsic to the nature of grains composed of monomers, in the same way as the single power law of the canonical parameterization is for compact grains.
\begin{figure}[htbp]
    \includegraphics[width=0.47\textwidth,left]{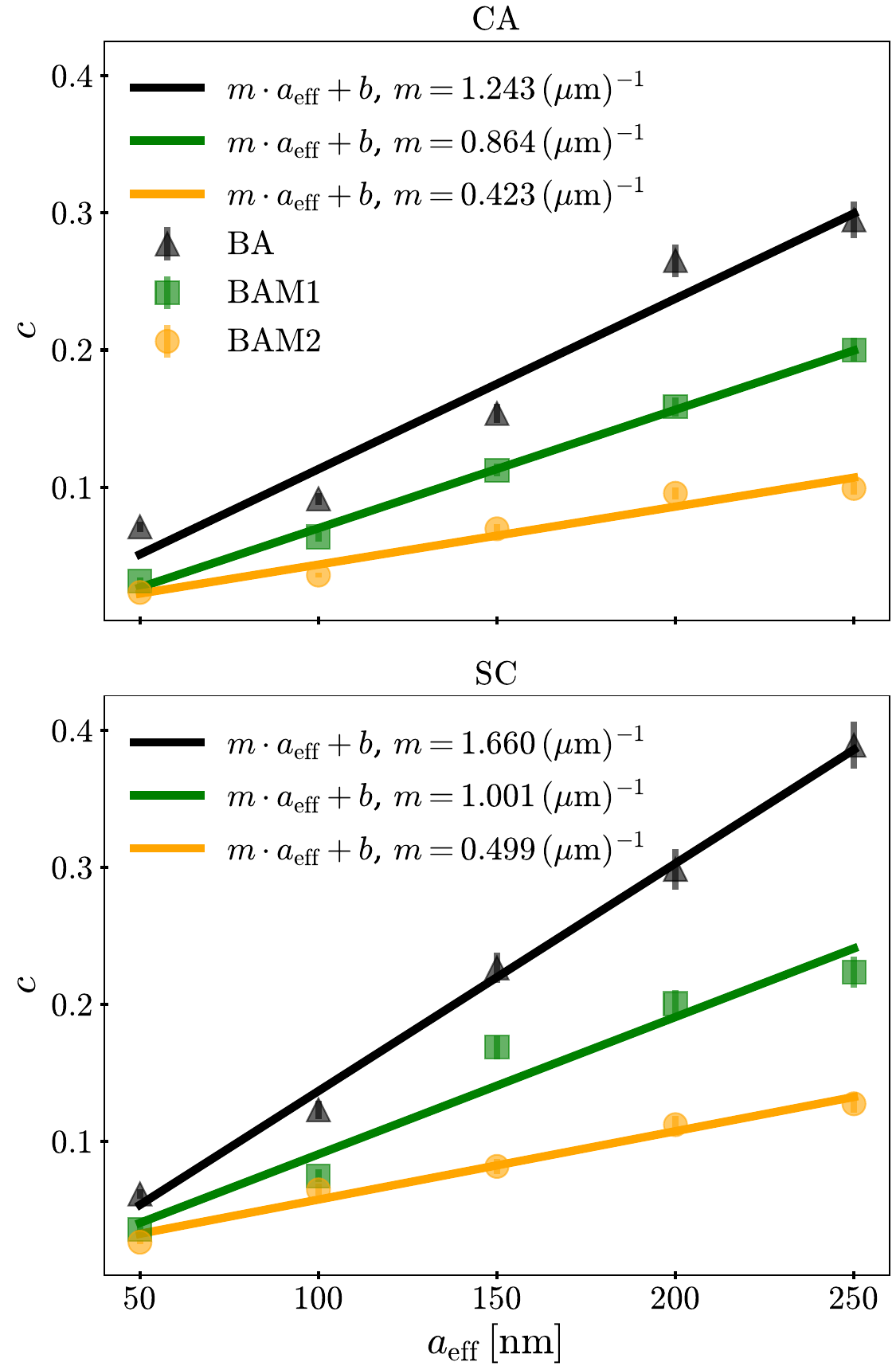}
	\caption{Constant part of the two-power-law parameterization $c$ vs. grain size for CA (left) and SC (right). The dotted lines show the linear regression for each grain model.}
	\label{fig:consts}
\end{figure}
It might be convenient to generalize the parameters of the parameterization and give them a functional form dependent on monomer density and grain size. Fig.~\ref{fig:consts} shows the constant part $c$ of the parameterization as a function of grain size for different monomer densities. The linear increase in $c$ with grain size has a different slope for each monomer density, and the slope roughly doubles from BAM2 to BAM1 and triples from BAM2 to BA. It is therefore possible that for each material there is a functional form of $c$ as a linear function, with the slope depending on the monomer density. To determine whether and how the exponents $q_1$ and $q_2$ depend on the material or monomer density requires a larger sample of monomer densities and refractive indices (materials).

We have shown that aggregated grains have a different functional form of $Q_\Gamma(\lambda)$ than compact grains, and they therefore do not reach angular velocities as high as compact grains would in the same circumstances, with differences of up to an order of magnitude for the smallest grains.
However, our results and the canonical parameterization agree for the least dense and largest grains (BA, $250\, \mathrm{nm}$). If the trend of increasing $c$ for a larger grain size and lower monomer density (see Fig.~\ref{fig:consts}) holds true, we would expect $\Gamma_{\mathrm{RAT}}$ to further increase for less dense and larger grains. The relation of a lower monomer density that leads to higher torques makes intuitive sense considering that it increases the total surface area of the grain and therefore the cross section for an interaction with radiation, just as increasing the grain size does. While the torque increases, $\omega_{\mathrm{RAT}}$ does not increase further for larger grains, however, and in some cases, it even decreases again for the largest grains. Aggregate grains larger than those simulated here may potentially not reach much higher angular velocities, as the drag also increases with size.\\
How much $\omega_{\mathrm{RAT}}$ increases for higher radiation energy densities strongly depends on the optical properties and therefore on the composition of the grains because different drag effects dominate in different environments. The significantly increased surface area and more complex geometry of aggregated grains has to be taken into account.\\
In environments with highly ionized gas and charged grains, such as an AGN, plasma drag in addition to gas drag and photon drag might additionally play a role for grain dynamics \citep{DraineLazarian1998}.  
However, the charging of BA dust adds an additional level of complexity, where individual monomers may carry different charges and even undergo some stochastic fluctuation over time \citep[see e.g.][]{Matthews2013,Ma2013,Matthews2018}. It is beyond the scope of this paper to include the plasma drag in the context of BA. However, the difference of the total grain charge of a BA compared to an equivalent sphere should be smaller than one order of magnitude, as shown in \cite{Ma2013} and \cite{Matthews2018}. Hence, we estimate that the canonical parameterization of RAT would still overestimate the resulting angular velocity $\omega_{\mathrm{RAT}}$ for ballistic aggregates in the proximity of an AGN.
The dust grains were assumed to have no relative velocity to the gas, but grains close to strong radiation fields might experience linear acceleration by the radiation pressure and reach high drift velocities relative to the surrounding gas. We emphasize that the resulting rotation of the BA grain would then also be influenced by mechanical torques \citep[METs, see e.g.][]{lazarian_subsonic_2007,Das2016, hoang_alignment_2018, reissl_mechanical_2023}, which work in a way similar to RATs, with gas collisions instead of photon collisions. However, it is still subject of debate how METs enhance the grain rotation in the presence of RATs \citep[see also][]{LazarianHoang2021ApJ}.

We emphasize once again that our modeling of RATs did not include the dynamical processes associated with an external magnetic field. As noted in LH07 for an analytical RAT model, the appearance of attractors in the $\Theta-\omega$ phase space is tightly connected to the ratio ${ q=\max(Q_{\mathrm{e1}})/\max(Q_{\mathrm{e2}}) }$ when an external field is present. This finding was later confirmed by \cite{herranen_radiative_2019} and \cite{Herranen2020} based on grain shapes that were modeled as Gaussian ellipsoids. We note, however, that a similar parameter $q$ for a pure mechanical alignment was not found \citep{Das2016,reissl_mechanical_2023}. Hence, it remains to be seen whether the ratio $q$ is a reliable proxy to predict the long-term stable alignment with the direction of the magnetic field, considering RATs as a spin-up mechanism of ballistic aggregates. Moreover, in the presence of a magnetic field, this is not only a matter of the efficiency of the RATs. Due to internal dissipation processes in paramagnetic and ferromagnetic materials, a fraction of the angular velocity $\omega$ is transferred into internal heat \citep[see e.g.][]{davis_polarization_1951,dolginov_orientation_1976,purcell_suprathermal_1979}. Consequently, the $\omega$ reported in our study may be lower, depending on the magnetic field strength and grain orientation. A similar effect of a reduction in $\omega$ for METs was shown by \cite{reissl_mechanical_2023}. However, to which degree a magnetic field would reduce the maximum possible angular velocity $\omega_{\mathrm{RAT}}$ of BAM aggregates remains to be addressed in forthcoming publications.

\section{Summary and outlook} \label{sect:summary}
We presented a numerical study of the impact of the grain shape on the radiative torque (RAT) efficiency and its dependence on the magnitude of the radiation field as well as the material and porosity of the dust grains. As a source of radiation, we considered O-type stars, an AGN, and the ISRF typical for the Milky Way. The dust grains were modeled by means of ballistic aggregation and migration (BAM) of monomers of various sizes. The aggregation process was biased to control the size and number of connections of the grain aggregates. We calculated the optical dust grain properties in the discrete dipole approximation using the DDSCAT code, considering the refractive indices of pure carbonaceous and silicate grains. Based on the optical properties of the dust aggregates and the radiation energy density, we calculated the maximum angular velocity $\omega_{\mathrm{RAT}}$ according to the RAT alignment theory. We note again that the conclusions presented here are still preliminary in nature, as a stable alignment of the grain rotation axis and the direction of incoming radiation was assumed and no magnetic fields were included. Our major findings are listed below.
\begin{itemize}
    \item We report that the ISRF produced average angular velocities $\omega_{\mathrm{RAT}}=10^6\,\mathrm{s}^{-1}$ for both carbon and silicate grains, whereas the O-type stars and AGN sources produced much larger $\omega_{\mathrm{RAT}}$ in a range of $10^8\,\mathrm{s}^{-1}$ to $10^{10}\,\mathrm{s}^{-1}$ for carbon grains and $10^7\,\mathrm{s}^{-1}$ to $10^{9}\,\mathrm{s}^{-1}$ for silicate grains. 
    \item Compared to the O-type stars, the angular velocity $\omega_{\mathrm{RAT}}$ of grains for the AGN did not increase in a similar manner because the photon emission drag forces on the grains increase. The reason is that the dust temperatures in the AGN environment are higher.    
    \item Considering an entire ensemble of BAM grain aggregates, we proposed a new parameterization for the RAT efficiency that follows a broken power law with a constant part for shorter wavelengths followed by two distinct power laws to cover longer wavelengths. The proposed parameterization describes the average angular velocities $\omega_{\mathrm{RAT}}$ reached by the ensembles of grains more precisely than the canonical parameterization derived for compact grains.
    \item According to our DDSCAT calculations, individual grain aggregates may have sign changes in the RAT efficiency over the entire wavelength regime. This reduces the net torque integrated over wavelength. However, the appearance of this sign changes affects the average grain spin-up behavior of the entire ensemble only marginally.
    \item The stronger stellar source and the AGN produce angular velocities that are potentially high enough for a rotational disruption of the grains. The results for the angular velocities obtained here could be used to study the rotational disruption of ballistic aggregate grains on the level of individual grains in the future.
\end{itemize}
Our results for a parameterization with two power laws for grains composed of monomers, potentially improved by expansion to a parameterization for the distribution of $\omega_{\mathrm{RAT}}$ values, can facilitate detailed studies of grain rotation including grain size, porosity, and material by replacing computationally expensive DDSCAT simulations of individual grains. Further improvements could be made by dropping the assumption of unobscured source radiation, including isotropic radiation, surface effects, $\mathrm{H}_2$ formation, and other additional drag effects. Especially in order to apply the results to grain alignment, a general orientation of the grains as well as magnetic fields should be added to complement the analysis.

\begin{acknowledgements}
We thank the anonymous referee for a constructive report. Special thanks go to Thiem Hoang for numerous enlightening discussions. R.S.K. also thanks the Harvard-Smithsonian Center for Astrophysics and the Radcliffe Institute for Advanced Studies for their hospitality during his sabbatical, and the 2024/25 Class of Radcliffe Fellows for highly interesting and stimulating discussions. S.R., J.A.J., and R.S.K. acknowledge financial support from the Heidelberg cluster of excellence (EXC 2181 - 390900948) “{\em STRUCTURES}: A unifying approach to emergent phenomena in the physical world, mathematics, and complex data”, specifically via the exploratory project EP 4.4. S.R. and R.S.K. also thank for support from Deutsche Forschungsgemeinschaft (DFG) via the Collaborative Research Center (SFB 881, Project-ID 138713538) 'The Milky Way System' (subprojects A01, A06, B01, B02, and B08). And we thank for funding form the European Research Council in the ERC synergy grant “{\em ECOGAL} – Understanding our Galactic ecosystem: From the disk of the Milky Way to the formation sites of stars and planets” (project ID 855130). The project made use of computing resources provided by {\em The L\"{a}nd} through bwHPC and by DFG through grant INST 35/1134-1 FUGG. Data are in part stored at SDS@hd supported by the Ministry of Science, Research and the Arts and by DFG through grant INST 35/1314-1 FUGG.
\end{acknowledgements}

\bibliographystyle{aa}
\bibliography{./bibtex.bib}

\appendix
\onecolumn
\section{Figures}\label{ap:figures}
In this section we present the full scope of our simulation results in Figs.\ref{fig:av_trq+2p_para}-\ref{fig:av_wrat+Tatsuuma_para}.

\noindent\begin{minipage}{\textwidth}
	\centering    \includegraphics[width=0.98\textwidth,left]{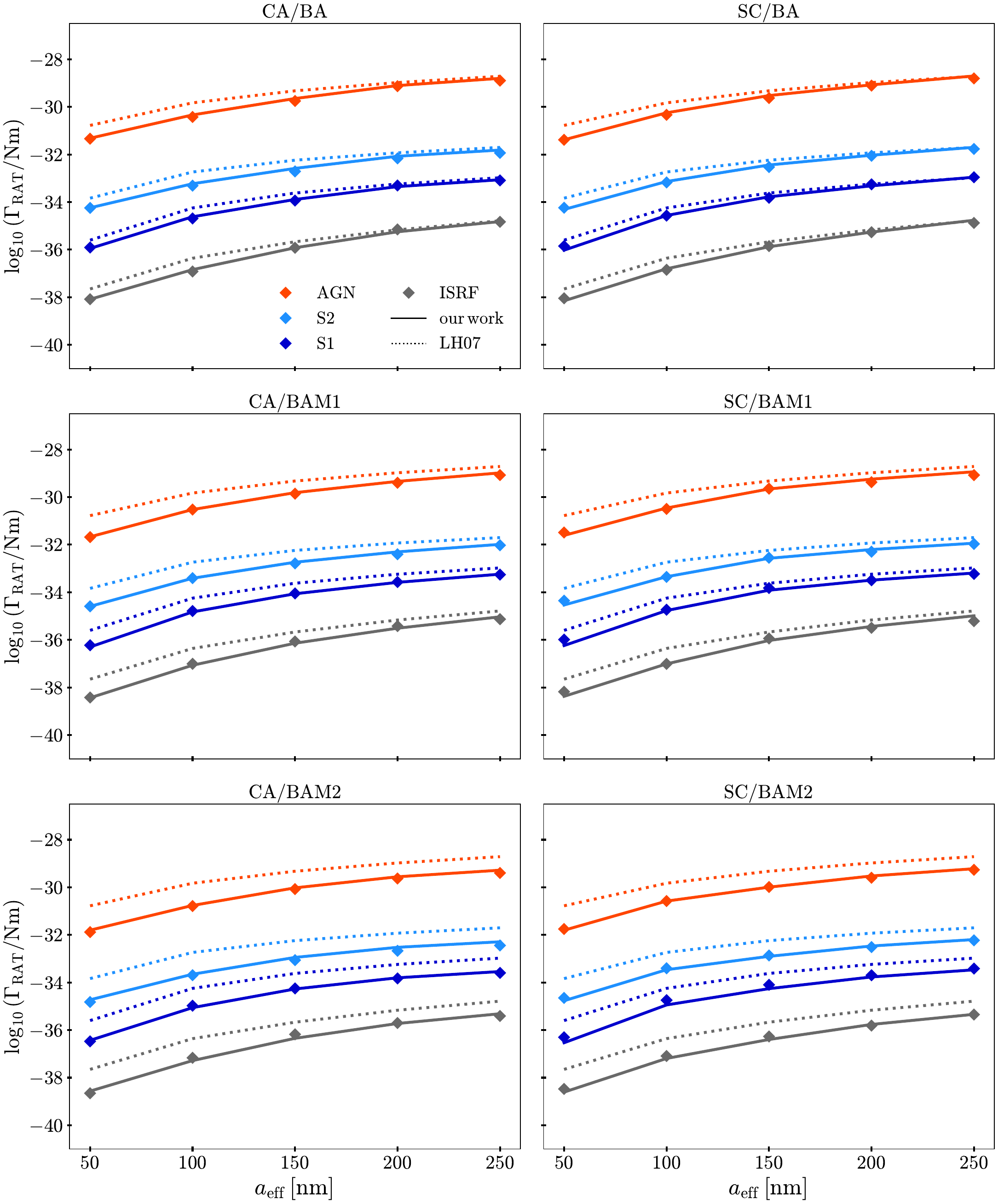}
	\captionof{figure}{RATs $\Gamma_{\mathrm{RAT}}$ (solid line) per grain size $a_{\mathrm{eff}}$ calculated with the two-power-law parametrization is shown in comparison to the RATs predicted by 
    the canonical parametrization suggested in LH07 (dotted line) as well as the average of the DDSCAT simulation data (diamonds). Each color represents one of the different sources i.e. ISRF (gray), S1 (dark blue), S2 (light blue), or AGN (red), respectively, utilizing the CA grain material (left column) and the SC material (right column) with monomer-densities BA (top row), BAM (middle row), or BAM2 (bottom row). This figure is to be compared with Fig.~\ref{fig:Hists}.}
	\label{fig:av_trq+2p_para}
\end{minipage}
\newpage

\begin{figure*}[htbp]
	\centering
	\includegraphics[width=\textwidth]{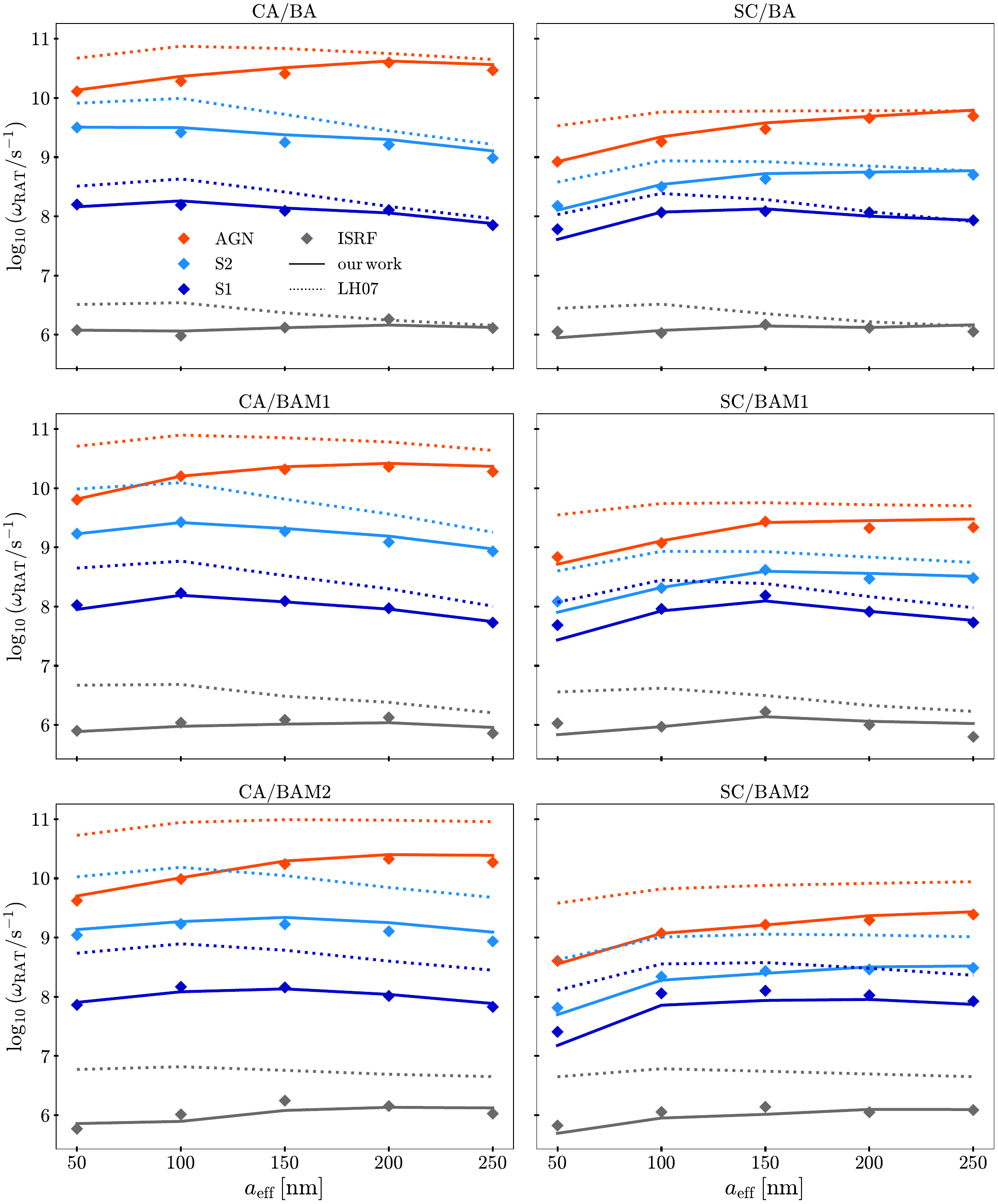}
	\caption{Same as Fig.~\ref{fig:av_trq+2p_para} but for angular velocity $\omega_{\mathrm{RAT}}$. We emphasize that the canonical parametrization may over-predict the magnitude of $\omega_{\mathrm{RAT}}$ up to one order of magnitude higher for BAM grain aggregates as the two-power-law parametrization.}
	\label{fig:av_wrat+2p_para}
\end{figure*}

\begin{figure*}[htbp]
	\centering
	\includegraphics[width=\textwidth]{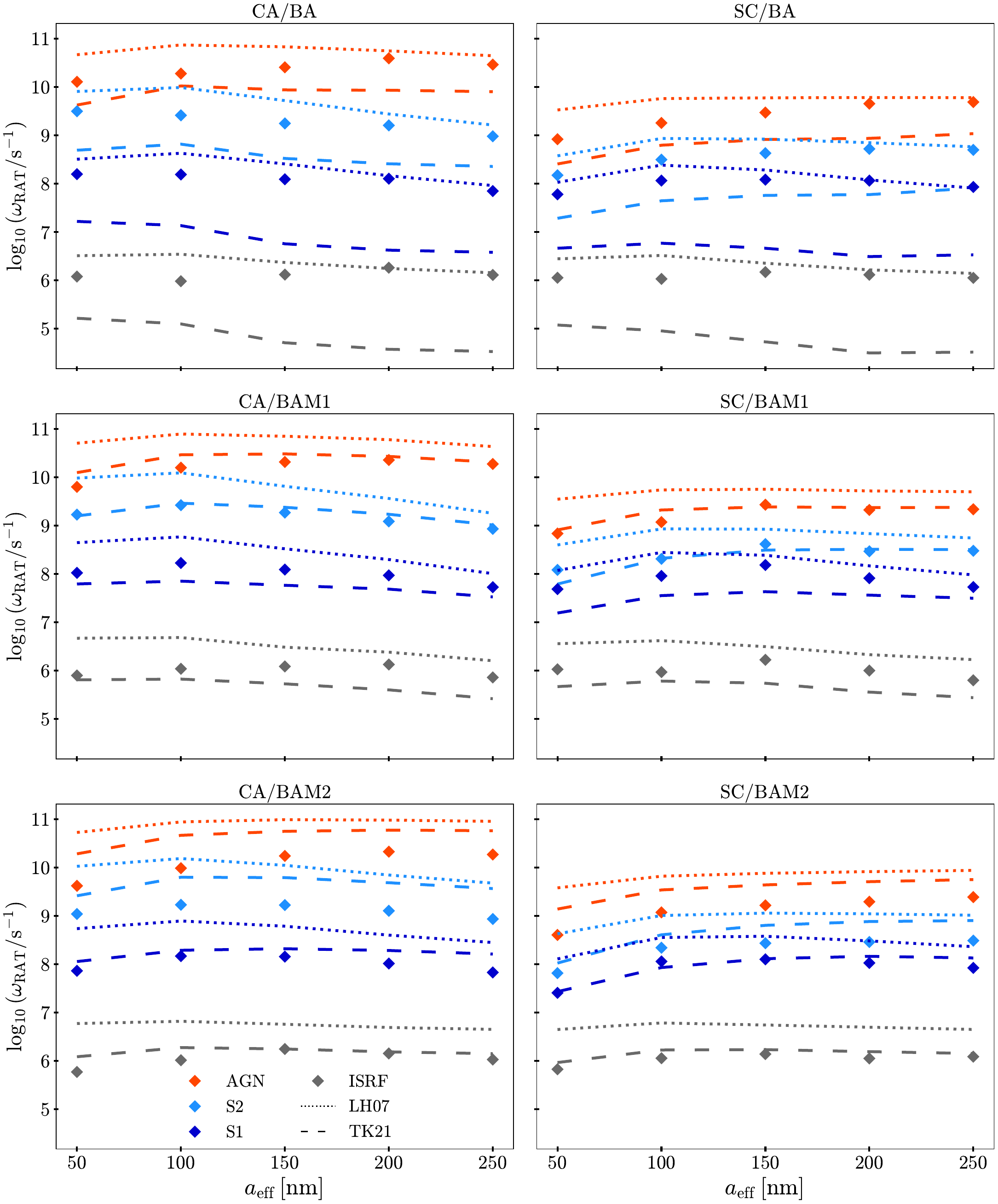}
	\caption{Same as Fig.~\ref{fig:av_wrat+2p_para} but showing the canonical parametrization of LH07 (dotted lines) in comparison with the parametrization for porous grains suggested in TK21 (dashed lines).}
	\label{fig:av_wrat+Tatsuuma_para}
\end{figure*}
\newpage
\FloatBarrier

\section{Two power-law parametrization fit parameters}
\label{ap:2pp_params}
For completeness we provide the full set of fitted parameters of the two-power-law RAT parametrization in table \ref{tab:params_CA} for CA grains and in table \ref{tab:params_SC} for SC grains.
\vspace{10mm}

\noindent\begin{minipage}{\textwidth}
	\captionof{table}{Fit parameters for CA}
	\label{tab:params_CA}
	\centering
	\begin{tabular}{r c c c c c c}
        \toprule
        monomer-density & $q_1$ & $q_2$ & $b$ & $a_{\mathrm{eff}}$ [nm] & $c$ & $a$ \\
        \midrule
        \multirow{5}*{BA} & \multirow{5}*{$-6.690 \pm 0.007$} & \multirow{5}*{$-1.0032 \pm 0.0003$} & \multirow{5}*{$14 \pm 5$} & 50 & $0.071 \pm 0.004$ & $3.202 \pm 0.026$ \\
        & & & & 100 & $0.091 \pm 0.004$ & $3.083 \pm 0.022$ \\
        & & & & 150 & $0.154 \pm 0.007$ & $2.853 \pm 0.021$ \\
        & & & & 200 & $0.265 \pm 0.012$ & $2.629 \pm 0.018$ \\
        & & & & 250 & $0.295 \pm 0.013$ & $2.588 \pm 0.018$ \\
        \midrule
        \multirow{5}*{BAM1} & \multirow{5}*{$-6.659 \pm 0.011$} & \multirow{5}*{$-1.02855 \pm 0.00022$} & \multirow{5}*{$12 \pm 4$} & 50 & $0.0315 \pm 0.0027$ & $3.21 \pm 0.04$ \\
        & & & & 100 & $0.064 \pm 0.003$ & $2.891 \pm 0.025$ \\
        & & & & 150 & $0.112 \pm 0.005$ & $2.655 \pm 0.018$ \\
        & & & & 200 & $0.159 \pm 0.007$ & $2.521 \pm 0.018$ \\
        & & & & 250 & $0.200 \pm 0.009$ & $2.434 \pm 0.017$ \\
        \midrule
        \multirow{5}*{BAM2} & \multirow{5}*{$-6.073 \pm 0.007$} & \multirow{5}*{$-1.01848 \pm 0.00018$} & \multirow{5}*{$13 \pm 5$} & 50 & $0.0232 \pm 0.0014$ & $3.13 \pm 0.03$ \\
        & & & & 100 & $0.0358 \pm 0.0016$ & $2.916 \pm 0.023$ \\
        & & & & 150 & $0.070 \pm 0.003$ & $2.614 \pm 0.022$ \\
        & & & & 200 & $0.096 \pm 0.004$ & $2.481 \pm 0.019$ \\
        & & & & 250 & $0.099 \pm 0.005$ & $2.467 \pm 0.019$ \\
        \bottomrule
	\end{tabular}
\end{minipage}
\vspace{5mm}

\noindent\begin{minipage}{\textwidth}
	\captionof{table}{Fit parameters for SC}
    \label{tab:params_SC}
	\centering
	\begin{tabular}{r c c c c c c}
        \toprule
        monomer-density & $q_1$ & $q_2$ & $b$ & $a_{\mathrm{eff}}$ [nm] & $c$ & $a$ \\
        \midrule
        \multirow{5}*{BA} & \multirow{5}*{$-6.168 \pm 0.008$} & \multirow{5}*{$-0.7434 \pm 0.0004$} & \multirow{5}*{$13 \pm 5$} & 50 & $0.062 \pm 0.004$ & $3.11 \pm 0.03$ \\
        & & & & 100 & $0.123 \pm 0.007$ & $2.785 \pm 0.026$ \\
        & & & & 150 & $0.227 \pm 0.011$ & $2.523 \pm 0.020$ \\
        & & & & 200 & $0.299 \pm 0.015$ & $2.412 \pm 0.020$ \\
        & & & & 250 & $0.389 \pm 0.017$ & $2.311 \pm 0.017$ \\
        \midrule
        \multirow{5}*{BAM1} & \multirow{5}*{$-6.393 \pm 0.007$} & \multirow{5}*{$-0.7033 \pm 0.0003$} & \multirow{5}*{$12 \pm 4$} & 50 & $0.0360 \pm 0.0022$ & $3.18 \pm 0.03$ \\
        & & & & 100 & $0.075 \pm 0.005$ & $2.832 \pm 0.029$ \\
        & & & & 150 & $0.169 \pm 0.009$ & $2.494 \pm 0.021$ \\
        & & & & 200 & $0.201 \pm 0.010$ & $2.428 \pm 0.020$ \\
        & & & & 250 & $0.224 \pm 0.011$ & $2.387 \pm 0.020$ \\
        \midrule
        \multirow{5}*{BAM2} & \multirow{5}*{$-5.386 \pm 0.008$} & \multirow{5}*{$-0.8477 \pm 0.0003$} & \multirow{5}*{$11 \pm 5$} & 50 & $0.0266 \pm 0.0014$ & $2.779 \pm 0.027$ \\
        & & & & 100 & $0.065 \pm 0.004$ & $2.36 \pm 0.03$ \\
        & & & & 150 & $0.082 \pm 0.005$ & $2.256 \pm 0.028$ \\
        & & & & 200 & $0.112 \pm 0.007$ & $2.127 \pm 0.023$ \\
        & & & & 250 & $0.128 \pm 0.006$ & $2.078 \pm 0.020$ \\
        \bottomrule
	\end{tabular}
\end{minipage}

\end{document}